\documentclass[mathpazo]{cicp}

\begin{document}

\title{Recovering the damping rates of cyclotron damped plasma waves from simulation data}

\author[Schreiner C et.~al.]{Cedric Schreiner\affil{1,2}\comma\corrauth,
      Patrick Kilian\affil{1}, and Felix Spanier\affil{1}}
\address{\affilnum{1}\ Centre for Space Research, 
	  North-West University, 
	  Potchefstroom 2520, South Africa. \\
          \affilnum{2}\ Lehrstuhl f\"ur Astronomie, 
	  Universit\"at W\"urzburg, 
	  97074 W\"urzburg, Germany.}
\emails{{\tt cschreiner@astro.uni-wuerzburg.de} (C.~Schreiner), {\tt mail@petschge.de} (P.~Kilian),
         {\tt felix@fspanier.de} (F.~Spanier)}

\begin{abstract}
Plasma waves with frequencies close to the particular gyrofrequencies of the charged particles in the plasma lose energy due to cyclotron damping.
We briefly discuss the gyro-resonance of low frequency plasma waves and ions particularly with regard to particle-in-cell (PiC) simulations.
A setup is outlined which uses artificially excited waves in the damped regime of the wave mode's dispersion relation to track the damping of the wave's electromagnetic fields.
Extracting the damping rate directly from the field data in real or Fourier space is an intricate and non-trivial task.
We therefore present a simple method of obtaining the damping rate $\Gamma$ from the simulation data.
This method is described in detail, focusing on a step-by-step explanation of the course of actions.
In a first application to a test simulation we find that the damping rates obtained from this simulation generally are in good agreement with theoretical predictions.
We then compare the results of one-, two- and three-dimensional simulation setups and simulations with different physical parameter sets.
\end{abstract}


\maketitle

\section{Introduction}
\label{sec:introduction}

Turbulence in a magnetized plasma, for example in the solar wind, develops a cascading spectrum of low frequency waves with ever shorter wave lengths.
The spectrum is limited by processes of wave damping, such as Landau damping or the cyclotron resonance for waves propagating perpendicular or parallel to the magnetic field (or by a mixture of both for oblique waves).
Since computer simulations of various plasma phenomena and especially of plasma turbulence are more and more common, it is also interesting to take a closer look at the representation of damping mechanisms in the simulation.

In this article we pick the cyclotron resonance of ions and low frequency waves, i.e. waves with a frequency $\omega$ below the cyclotron frequency $\Omega$ of the resonating particles.
This process can be easily modeled using waves propagating parallel to a background magnetic field and a thermal spectrum of protons.
We choose the particle-in-cell (PiC) approach, because it is a self-consistent method which treats kinetic effects in the plasma.
Thus, it is expected that a PiC simulation captures cyclotron damping correctly.
Of course, PiC is not the only numerical approach which includes cyclotron damping and other types of code might be used as well to study wave damping.

Determining damping (or growth) rates of plasma waves is not a trivial task.
Properties, such as the wave length or wave number, can easily be obtained from looking at a real space representation of the field data or its Fourier transform in space.
Another Fourier transform in time yields frequency information and the dispersion relation of the whole wave mode.
It is even possible to recover the polarization of single waves or whole wave modes by adequately combining different components of the electromagnetic fields.
However, obtaining the damping rate directly from real or Fourier space electromagnetic field data is challenging.

One approach, though, is to resolve the dispersion relation of the wave mode in question to such an extend that a broadening in frequency can be observed.
For a single wave with frequency $\omega_0$ it may be assumed that the wave's intensity, represented by its energy density $W$, follows a Lorentz profile over frequency $\omega$:
\begin{equation}
	W(\omega) = \frac{W_0 \, \Gamma^2}{(\omega - \omega_0)^2 + \Gamma^2}
	\label{eq:lorentz}
\end{equation}
which is centered around the wave's frequency $\omega_0$ and has a width at half maximum of $2 \, \Gamma$, where $\Gamma$ is the damping rate.
Thus, fitting the Lorentz profile from Eq. (\ref{eq:lorentz}) to the data yields the damping rate of the wave.
This is a tedious process and the precision of the results strongly depends on a high resolution of the frequencies in the dispersion relation, which is often only achieved by the use of a massive amount of computational resources.

A simple and fast possibility of studying wave damping (or any other interaction of waves and particles) is to analyze the composition of the total energy in the simulation.
By comparing the development of the total field energy and the kinetic energy of the particles -- quantities which are often computed during the simulation and saved for diagnostic purposes -- it becomes obvious when and to which extent energy is transferred between waves and particles.
However, no information about the wave's properties, such as frequency and wave number, are contained in such a study and several similar processes cannot be distinguished.
It might even not be possible to tell which wave mode participates in the process, especially in fully kinetic simulations which might contain several possible candidates.

A method to determine the growth rates of different wave modes has been proposed by Koen et al. \cite{koen_2012}, who study the interaction of three populations of electrons (cold, warm and beam) and electron plasma, acoustic, and beam modes in an electrostatic PiC simulation.
Their method is based on the Fourier transformation of field data both in time and space which is then used to obtain the energy density $W(k, \, \omega)$ as a function of the wave number $k$ and the frequency $\omega$.
Thus it is possible to discriminate between different waves or wave modes (described by their wave numbers $k$ and corresponding frequencies $\omega$).
Whereas Koen et al. \cite{koen_2012} analyze several wave modes at once and thus need a high resolution in $\omega$ to be able to separate the different modes, we will show that such a high resolution is not necessary for their method to work.

In the article at hand, we focus on a step-by-step explanation of the method of \cite{koen_2012} and its extension to cyclotron damping.
We measure the damping rates of individual low-frequency L-mode waves undergoing cyclotron damping in an electromagnetic PiC simulation.
On this example we demonstrate that the method can even be applied if the spectral resolution is low and the frequency range of the wave mode in question is not resolved.
It is sufficient to distinguish these waves only by their wave number.
Furthermore we apply the method to a series of simulations to investigate the quality of the representation of cyclotron damping in PiC simulations in different numerical and physical scenarios.

The article is organized as follows:
We first give a brief description of the numerical methods and the setup of our PiC simulations in Sects. \ref{sec:methods} and \ref{sec:simulation}.
The approach to measuring the damping rate is presented in Sect. \ref{sec:damping_rate} and illustrated by its application to an example simulation.
Studies of the effects of the simulation's dimensionality, i.e. a one-, two- or three-dimensional setup, and different sets of physical parameters are discussed in Sects. \ref{sec:dimensions} and \ref{sec:parameters}, respectively.
Finally, we summarize our results in Sect. \ref{sec:discussion}.

\section{Numerical methods and theory}
\label{sec:methods}

\subsection{PiC code}
\label{sec:pic}

Our PiC simulations are carried out using the \emph{ACRONYM} code \cite{kilian_2012}, which is designed to employ second-order numerical schemes throughout the code.
The \emph{ACRONYM} code is an electromagnetic, fully relativistic, explicit PiC code which supports one-, two- or three-dimensional setups, while electromagnetic fields and particle velocities are always treated as three-dimensional vectors.

Furthermore we use the initial conditions described by \cite{schreiner_2014_a} to excite one or more waves at the beginning of the simulation.
The excitation mechanism allows for the creation of damped waves in the initialization phase.
These waves are then damped during the simulation (``free decay''), losing field energy to the particles.
The energy loss can be measured and is the basis for our analysis of the damping rate.

\subsection{Warm plasma dispersion relation}
\label{sec:warm_plasma}

For later comparison of simulation results and theory we also compute the theoretical damping rate from the warm plasma dispersion relation.
We follow the notation of \cite{chen_2013}, who give the generalized dispersion relation $D(\omega_c, k_\parallel)$ for the parallel propagating, left-handed L-mode \cite{kennel_1966} in their Eq. (1):
\begin{equation}
	0 = D(\omega_c, k_\parallel) = \omega_c^2 - k_\parallel^2 \, c^2 + \sum_\mathrm{s} \, \omega^2_\mathrm{p,s} \, I_\mathrm{s}.
	\label{eq:warm_disp}
\end{equation}
Here, $\omega_c = \omega + \imath \Gamma$ is the complex frequency, containing the real frequency $\omega$ and the growth / damping rate $\Gamma$ of a parallel propagating wave with wave number $k_\parallel$.
The index s denotes a particle species (such as electrons or protons), $\omega_\mathrm{p,s}$ is the plasma frequency of species s and $c$ is the speed of light.
Chen et al. \cite{chen_2013} then give a function $I_\mathrm{s}$ for a Maxwellian particle distribution (with potentially different plasma temperatures parallel and perpendicular to the background magnetic field):
\begin{equation}
	I_\mathrm{s} = \frac{\alpha_{\perp\,\mathrm{s}}^2}{\alpha_{\parallel\,\mathrm{s}}^2} - 1 + \left( \frac{\alpha_{\perp\,\mathrm{s}}^2}{\alpha_{\parallel\,\mathrm{s}}^2} \, \frac{\omega_c \mp \Omega_\mathrm{s}}{\pm \Omega_\mathrm{s}} + 1 \right) \, \frac{\pm \Omega_\mathrm{s}}{k_\parallel \, \alpha_{\parallel \, \mathrm{s}}} \, Z(\zeta_\mathrm{s}),
	\label{eq:warm_disp_maxwell}
\end{equation}
where $Z(\zeta_\mathrm{s})$ is the plasma dispersion function \cite{fried_1961} with the argument $\zeta_\mathrm{s} = (\omega_c \mp \Omega_\mathrm{s}) / (k_\parallel \, \alpha_{\parallel \, \mathrm{s}})$.
The cyclotron frequency $\Omega_\mathrm{s} = (q_\mathrm{s} \, B_0) / (m_\mathrm{s} \, c)$ may have a different sign for different particle species, depending on their charge $q_\mathrm{s}$.
The background magnetic field is denoted by $B_0$ and the particle mass by $m_\mathrm{s}$.
Normalized thermal velocities in parallel and perpendicular direction are given by the parameters $\alpha_{\parallel \, \mathrm{s}}$ and $\alpha_{\perp \, \mathrm{s}}$, which are defined via the temperature $T_{\parallel / \perp} = (m \, \alpha^2_{\parallel / \perp}) / 2$ (indices for particle species are omitted here).

Our Eq. (\ref{eq:warm_disp_maxwell}) is a slightly modified version of Eq. (13) from \cite{chen_2013}, where we have included alternating signs ($\pm$).
The upper signs are identical to the ones in the original equation and refer to the L-mode, whereas the lower signs refer to the right-handed, circularly polarized R-mode.
Thus, Eq. (\ref{eq:warm_disp}) describes both L- and R- mode, depending on the signs chosen in Eq. (\ref{eq:warm_disp_maxwell}).

The combination of Eqs. (\ref{eq:warm_disp}) and (\ref{eq:warm_disp_maxwell}) can be solved numerically and yields the complete dispersion relation for the L- or the R-mode in a magnetized, warm plasma.
Since these wave modes include only parallel propagating waves, the damping rate $\Gamma$ obtained from the dispersion relation refers to cyclotron damping and can be used to evaluate the results of our PiC simulations.

\section{Simulation setup for example simulations}
\label{sec:simulation}

In this section we describe a simple setup which can be used to analyze cyclotron damping of parallel propagating plasma waves.
The setup described is not meant to mirror a specific real-world counterpart, but is solely chosen for its numerical convenience.

Please note that wave numbers are discretized in a PiC simulation due to the discretization of physical space into individual grid cells.
Thus, each wave mode consists of a limited number of individual waves, which each can be characterized by their specific wave number $k$ and a corresponding frequency.
In the following, we refer to an individual oscillation with a wave number $k$ and frequency $\omega$ as ``a wave''.
Each wave mode in a simulation can be decomposed into a countable number of individual waves.
With the term ``waves of the L-mode'' to refer to the complete set of individual waves which make up the wave mode.

\subsection{Basic setup}
\label{sec:setup}

In order to obtain the best resolution of the wave and still minimize the computing time we choose a long but narrow three-dimensional simulation box with periodic boundary conditions.
The background magnetic field is chosen to point in the direction of the long edge of the simulation box.
The box length in the long (parallel) direction defines the maximum wave length for any parallel propagating wave and has to be chosen according to the desired specifications of the excited waves.
The radius of a particle's undisturbed gyration about the background magnetic field is usually a good indication for the length of the two short (perpendicular) edges of the box.
Having the short edges be at least twice as long as the Larmor radius of a particle traveling with thermal speed allows to cover the full gyration of most particles.

The plasma is characterized by the use of a few parameters:
We set the background magnetic field $B_0$, which fixes the cyclotron frequencies $\Omega_\mathrm{s}$ of the particles.
The plasma frequency of the electrons $\omega_\mathrm{p,e}$ determines the electron density $n_\mathrm{e}$.
Since we only study electron-proton plasmas, the density of the protons is equal to $n_\mathrm{e}$ and the plasma frequency of the protons is then $\omega_\mathrm{p,p} = \omega_\mathrm{p,e} \, \sqrt{m_\mathrm{e}/m_\mathrm{p}}$, with the masses of electrons and protons $m_\mathrm{e}$ and $m_\mathrm{p}$.
Similarly, the temperature $T$ of the (Maxwellian) plasma is set by selecting a thermal speed of the electrons $v_{\mathrm{th,e}} = \sqrt{k_B \, T_\mathrm{e} / m_\mathrm{e}}$, where $k_B$ is the Boltzmann constant.
In the simple case of isotropic temperature ($T_\parallel \! = \! T_\perp \! = \! T$), this translates to $\alpha = 2 \, v_\mathrm{th}^2 \, / \, k_B$ (once again omitting indices for particle species).
With $T_\mathrm{e} \! = \! T_\mathrm{p} \! = \! T$, the thermal speed of the protons is $v_{\mathrm{th,p}} = v_{\mathrm{th,e}} \, \sqrt{m_\mathrm{e}/m_\mathrm{p}}$.
Finally, we choose a mass ratio $m_\mathrm{p}/m_\mathrm{e}$, where the electron mass is kept at its natural value, whereas the proton mass can be changed.
To reduce computing time a lower, artificial mass ratio is desirable if proton effects, such as cyclotron damping of L-mode waves, are to be analyzed.

Physical and numerical parameters used for the example simulation presented in this article can be found in Tables \ref{tab:setup_phys} and \ref{tab:setup_num}.
Additional parameters to the ones discussed above are the amplitude of the excited waves, $\delta B$, the grid spacing and the length of the time step, $\Delta x$ and $\Delta t$, and the number of time steps, $N_\mathrm{t}$.
The size of the simulation box is given in cells, with $N_\parallel$ and $N_\perp$ characterizing the directions parallel and perpendicular to the background magnetic field $\vec{B_0}$.

\begin{table}[h]
	\caption{Physical parameters for an example simulation.}
	\label{tab:setup_phys}
	\centering
	\begin{tabular}{c c c c}
		\hline
		\noalign{\smallskip}
		$\omega_\mathrm{p,e} \, (\mathrm{rad ~ s}^{-1})$ & $|\Omega_\mathrm{e}| \, (\omega_\mathrm{p,e})$ & $v_\mathrm{th,e} \, (c)$ & $\delta B \, (B_0)$ \\
		\noalign{\smallskip}
		\hline
		\noalign{\smallskip}
		$2.00 \cdot 10^8$ & $4.40 \cdot 10^{-1}$ & $5.00 \cdot 10^{-2}$ & $5.00 \cdot 10^{-3}$ \\
		\noalign{\smallskip}
		\hline
	\end{tabular}
\end{table}

\begin{table}[h]
	\caption{Numerical parameters for an example simulation.}
	\label{tab:setup_num}
	\centering
	\begin{tabular}{c c c c c c}
		\hline
		\noalign{\smallskip}
		$N_\parallel \, (\Delta x)$ & $N_\perp \, (\Delta x)$ & $N_t \, (\Delta t)$ & $\Delta x \, (c \, \omega_\mathrm{p,e}^{-1})$ & $\Delta t \, (\omega_\mathrm{p,e}^{-1})$ & $m_\mathrm{p} \, (m_\mathrm{e})$ \\
		\noalign{\smallskip}
		\hline
		\noalign{\smallskip}
		$1200$ & $64$ & $50000$ & $3.53 \cdot 10^{-2}$ & $2.04 \cdot 10^{-2}$ & $40.0$ \\
		\noalign{\smallskip}
		\hline
	\end{tabular}
\end{table}

With the start of the simulation, the random thermal motion of the particles creates electromagnetic fluctuations and eventually all kinds of physically allowed plasma waves of different wave modes.
Waves in the damped regime of different wave modes are hardly produced, since any energy in the electromagnetic fields of such a wave is eventually dissipated and transferred to the particles.
However, damped waves can still be studied if a decent amount of energy is artificially deposited in the electromagnetic fields of this wave.
For doing so, we employ the initialization mechanism outlined in \cite{schreiner_2014_a} which excites one or more specific waves at the start of the simulation.

Using this method, wave excitation is achieved in three steps:
First, the frequency of a wave with wave number $k_\parallel$ is computed using the cold plasma dispersion relation (e.g. \cite{stix_1962}):
\begin{equation}
	|k_\parallel| = \frac{\omega}{c} \sqrt{1 - \frac{\omega_\mathrm{p}^2}{(\omega \pm \Omega_\mathrm{e}) (\omega \pm \Omega_\mathrm{p})}},
	\label{eq:cold_disp}
\end{equation}
where $\omega_\mathrm{p} = \sqrt{\omega_\mathrm{p,e}^2 + \omega_\mathrm{p,p} ^2}$ is the plasma frequency of the electron-proton plasma and the different signs discriminate between R- and L-mode (upper or lower sign, respectively).
Next, the polarization and amplitude of the wave are fixed by setting either the electric or magnetic field and then deriving the remaining one via Maxwell's equations.
For L- and R-mode waves, the fields are arranged as transverse, circularly polarized, plane waves with frequency $\omega$ and wave number $k_\parallel$.
Last, the electromagnetic fields are deployed on the grid of the PiC simulation.
Still during the initialization, the Lorentz force at each particle's position is computed to push the particles accordingly and obtain a consistent picture of the wave (see \cite{schreiner_2014_a}, Eqs. (6) and (7) for equations of motion).

\subsection{Effects of finite plasma temperature}
\label{sec:warm_plasma_effects}

\begin{figure}[t!]
	\centering
	\includegraphics[width=0.6\linewidth]{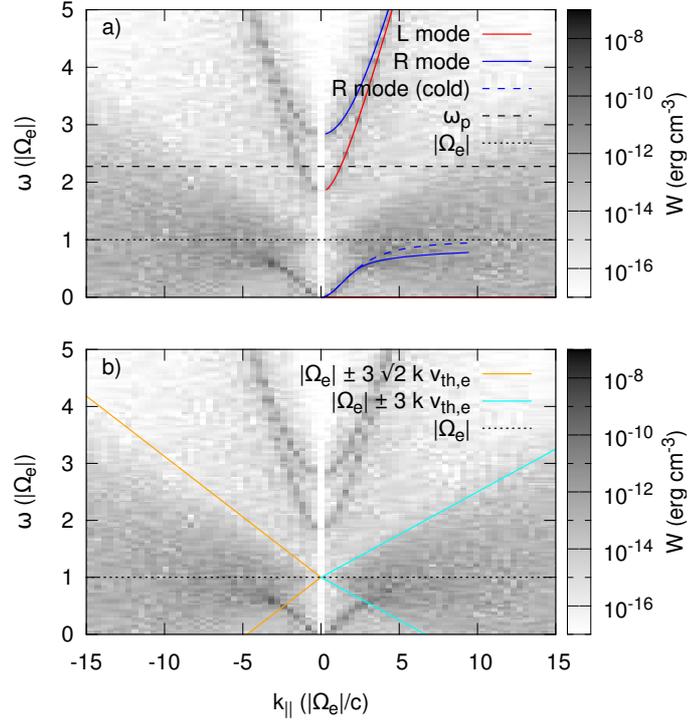}
	\caption{Dispersion plot obtained from a transverse magnetic field component. The color coded data is the same in both panels and shows the L- and the R-mode. The dark spot at \mbox{$k_\parallel \approx -3 \, c \, |\Omega_\mathrm{e}|^{-1}$} represents an initially excited wave in the dispersive regime of the R-mode. Solid lines in panel a) are the correct dispersion relations in a warm plasma, the dashed line shows the result from cold plasma theory. The divergence of the correct (warm) and the approximated (cold) theory becomes obvious as the curves approach $|\Omega_\mathrm{e}|$.
		The darker regions around $k_\parallel = 0$, $\omega = |\Omega_\mathrm{e}|$ which stretch out to larger $k_\parallel$ can be identified as the regime in which cyclotron damping becomes important. Panel b) shows the theoretical boundaries of this regime from Eq. (\ref{eq:res_r}) (left side) and the result of a modified version of this equation (right side), which seems to be more appropriate.}
	\label{fig:excitation}
\end{figure}

A point of criticism might be that the dispersion relation (\ref{eq:cold_disp}) is derived in the cold plasma limit, whereas cyclotron damping is a thermal effect.
However, the excited wave ``finds'' the correct place along the warm plasma dispersion relation, which naturally evolves in the PiC simulation, as Fig. \ref{fig:excitation} shows.
Using the cold plasma dispersion relation for the initial calculations in the code therefore does not lead to further problems or inconsistencies, especially for waves with small amplitudes.

To test whether a wave is affected by thermal effects and damping, a resonance condition can be used.
For example, Gary \& Nishimura \cite{gary_2004} give a resonance condition for Landau and cyclotron resonance in their Eqs. (1a) and (1b).
For cyclotron damping in a Maxwellian plasma their resonance condition translates to:
\begin{align}
	\omega_\mathrm{res,R} &\ge |\Omega_\mathrm{e}| - 3 \,\sqrt{2} \, k_\parallel \, v_{\mathrm{th,e}},
	\label{eq:res_r}
	\\
	\omega_\mathrm{res,L} &\ge \Omega_\mathrm{p} - 3 \, \sqrt{2} \, k_\parallel \, v_{\mathrm{th,p}}
	\label{eq:res_l}
\end{align}
for the R- and L-mode, respectively.
In fact, our simulation results suggest similar conditions -- lacking only the factor of $\sqrt{2}$ (see Fig. \ref{fig:excitation}).

\section{Obtaining the damping rate}
\label{sec:damping_rate}

In order to discuss our method of obtaining damping rates from simulation data we set up two simulations using the parameters given in Tables \ref{tab:setup_phys} and \ref{tab:setup_num} in Sect. \ref{sec:setup}.
We choose to excite only waves with odd numerical wave number $k_\mathrm{num} = (k_\parallel \, N_\parallel \, \Delta x) / (2 \, \pi)$ in one of the simulations, and only waves with even $k_\mathrm{num}$ in the other.
This is, of course, not necessary, but it nicely illustrates the difference between $k_\mathrm{num}$ with and without excited wave modes in the dispersion plots (see Fig. \ref{fig:example_disp}).

To distinguish between a wave's physical frequency $\omega$ and its apparent frequency in the dispersion plots obtained from the simulation data, we introduce $\omega_\mathrm{num}$.
As with the numerical wave number $k_\mathrm{num}$, $\omega_\mathrm{num}$ is the wave's frequency measured in pixels or bins in the dispersion plot.
Most importantly, for low-frequency waves, whose physical frequency $\omega$ cannot be resolved due to insufficient spectral resolution of the dispersion plot, the numerical frequency will be $\omega_\mathrm{num} = 0$.
However, it is worth noting that the wave still propagates (very slowly) during the simulation and that $\omega_\mathrm{num} = 0$ does not necessarily imply $|\Gamma| > \omega$.

\subsection{Dispersion plots}
\label{sec:dispersion_plots}

To measure the damping rate of waves on the low frequency branch of the L-mode (or ion cyclotron waves) we examine the dispersion plots obtained from our simulations.
As seen in the top panel of Fig. \ref{fig:excitation}, the low frequency regime is usually under-resolved.
The lowest frequency in the dispersion plot is inversely proportional to the total run-time of the simulation.
To reduce the total number of time steps, the mass ratio is reduced to increase the proton cyclotron frequency and thus raise the frequencies of all waves on the low frequency branch of the L-mode.
However, this does not mean that the branch will be resolved in the dispersion plots.
Since we are interested in initially excited, but damped waves, the resolution cannot be improved by increasing the number of time steps, if the wave in question has already been dissipated until the end of the simulation.

\begin{figure}[h!]
	\centering
	\includegraphics[width=0.6\linewidth]{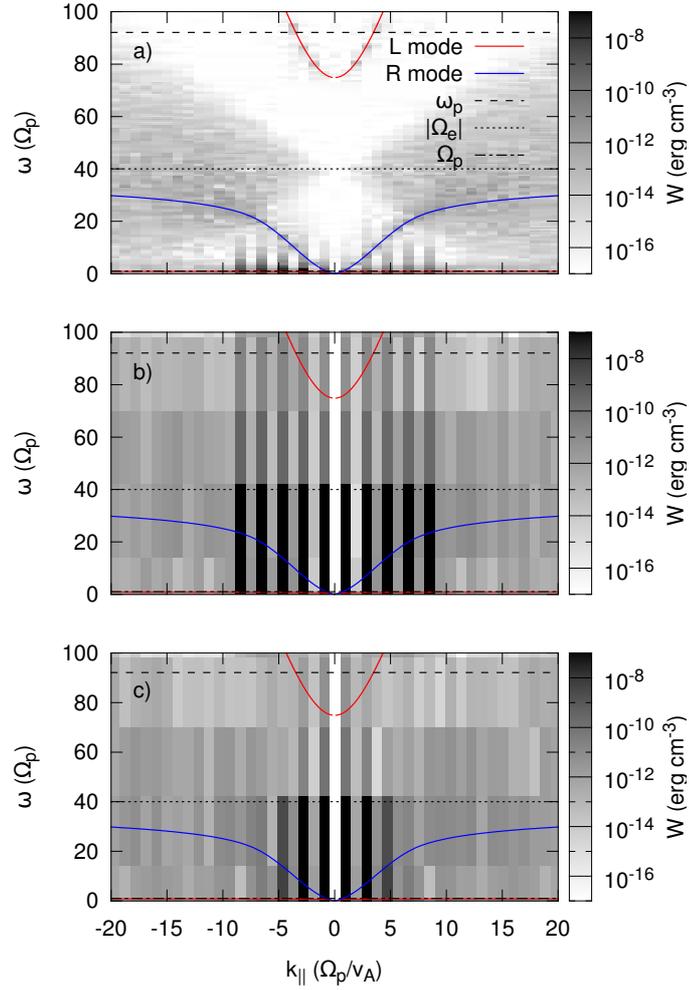}
	\caption{Dispersion relations from one component of the transverse magnetic field along the direction of the background magnetic field. The dispersion relation in panel a) contains data from the whole simulation, whereas the lower panels contain data from a shorter interval at an earlier (b) or later (c) stage of the simulation. Five waves have been excited at positive $k_\parallel$ (dark spots). While wave modes are resolved in a), no single mode can be seen in b) and c). However, the information about the waves' amplitudes is still conserved and can be read out. Note that the excited wave modes are symmetric in $k$ because the frequency is under-resolved and information about wave propagation is lost. Also note that the broadening of the excited waves in $\omega$ has no physical implications, but is mainly an artifact from using windowing during the Fourier transformation of the data.}
	\label{fig:example_disp}
\end{figure}

Even if the frequency of the waves can be resolved, there is probably no chance to see line broadening and to determine the damping rate using a Lorentz profile (\ref{eq:lorentz}) as a fit to the energy density $w(\omega)$ as a function of frequency.
However, with the method presented in this article, we are able to measure the damping rate even if the wave is not resolved in the dispersion plot, because although the information about the wave's frequency is lost, the information about its amplitude remains.
This information is stored in the lowest frequency bin ($\omega_\mathrm{num} = 0$), but at the correct wave number $k$ and can thus be easily accessed.
By measuring the amplitude (or energy density $W$) of the wave at multiple points in time throughout the simulation we are able to model $W(t)$ and obtain a damping rate.

The exact procedure is described in the following and illustrated by Fig. \ref{fig:example_disp}.
The simulation is split into several intervals with constant length $t_\mathrm{int}$ (measured in numerical units $\Delta t$).
The data from a specific interval $i \in \{1, 2, \ldots, N_t / t_\mathrm{int}\}$ can be used to produce dispersion relations which characterize the energy distribution in $k$-$\omega$-space during this interval.

In Fig. \ref{fig:example_disp} we present dispersion relations for one transverse component of the magnetic field along the direction of the background magnetic field $\vec{B_0}$.
Panel a) shows the dispersion relation obtained from the data of the entire simulation (i.e. all time steps), whereas the dispersion relations in panels b) and c) were produced at different intervals in time using only a subset $t_\mathrm{int}$ of time steps.
Compared to panel a) the resolution is considerably worse in panels b) and c), but in exchange differences in the intensities of the excited waves can be observed in the latter panels, representing time evolution.
Note that Fig. \ref{fig:example_disp} shows only excerpts of the complete dispersion plot, since only the region at small wave numbers and frequencies is relevant.
Also note that the total energy density $W_\mathrm{total}$ (i.e. the sum of the energy densities in all pixels) in the plots is not the same, since panel a) yields an average energy density over the course of the whole simulated time, whereas panels b) and c) represent averages over shorter intervals early or late in the simulation.
Due to the transfer of field energy to the particles, the total energy density $W_\mathrm{total,b}$ in Fig. \ref{fig:example_disp} b) exceeds $W_\mathrm{total,c}$ in \ref{fig:example_disp} c).
The average over the whole simulation, $W_\mathrm{total,a}$, lies somewhere between $W_\mathrm{total,b}$ and $W_\mathrm{total,c}$.

In the case of cyclotron damping, the dispersion relations for each transverse component of the electric and magnetic fields along the direction of $\vec{B_0}$ have to be computed.
Accumulating the energy density at a specific position $(k_\mathrm{num},\omega_\mathrm{num})$ in each dispersion relation gives the total energy density $W(t_i)$ of the respective wave during interval $i$.
Note that $\omega_\mathrm{num}$ is probably zero, since the frequency is not properly resolved and the wave in question can be entirely characterized by its wave number $k$.
Measuring $W$ in each interval yields $N_t / t_\mathrm{int}$ samples which represent $W(t)$.

\subsection{Energy density and damping rate}
\label{sec:energy_density}

\begin{figure}[ht]
	\centering
	\includegraphics[width=0.6\linewidth]{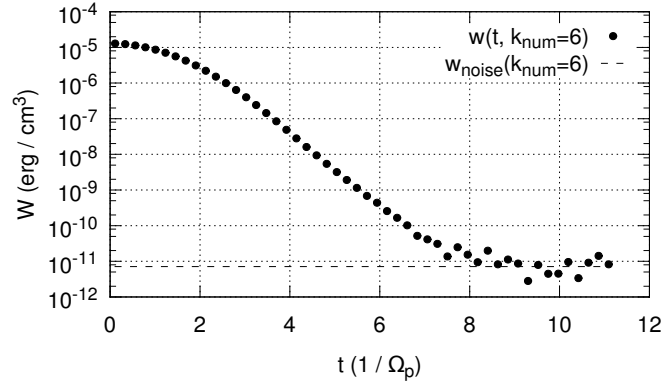}
	\caption{Energy density $W$ over time for a single wave with $k_\mathrm{num} = (k_{\parallel} \, L_\parallel) / (2\,\pi) = 6$, where $L_\parallel$ is the length of the simulation box parallel to the background magnetic field. The dots represent energy measurements from dispersion relations such as Fig. \ref{fig:example_disp} b) and c). Three intervals of the wave's evolution can be distinguished: a slow decay until $t \, \Omega_\mathrm{p} \approx 3$, exponential decay until $t \, \Omega_\mathrm{p} \approx 7$ and random fluctuations after the wave's energy is below the noise limit. The dashed line is the averaged energy density at $k_\mathrm{num} = 6$ in a simulation without wave excitation and represents the expected energy level of the background noise.}
	\label{fig:example_evolution}
\end{figure}

Summing up the energy densities of all field components for each wave in each interval yields the time evolution of the waves' energy densities (see Fig. \ref{fig:example_evolution}).
We expect the amplitude $A$ of a wave to decay as
\begin{equation}
	A(t) \propto \exp{(\Gamma \, t)}
	\label{eq:amplitude_decay}
\end{equation}
and thus the energy density to decay as
\begin{equation}
	W(t) \propto A(t)^2 \propto \exp{(2 \, \Gamma \, t)},
	\label{eq:energy_decay}
\end{equation}
where $\Gamma$ is the damping rate.

However, as can be seen from Fig. \ref{fig:example_evolution}, the energy density does not exhibit the expected behavior over the course of the whole simulation.
An initial phase exists, during which the decay proceeds more slowly than expected (i.e. not exponential).
The reason for this non-exponential onset might lie in the peculiarities of the initialization of the background plasma (excitation of plasma waves by the random motion of thermal particles) or in the processes following the excitation of a cold plasma wave which then lead to the establishment of the correct wave in a warm plasma.
This initial phase then transits to the exponential decay, which finally comes to a halt when the energy level of the thermal noise in the background plasma is reached.

\begin{figure}[t]
	\centering
	\includegraphics[width=0.6\linewidth]{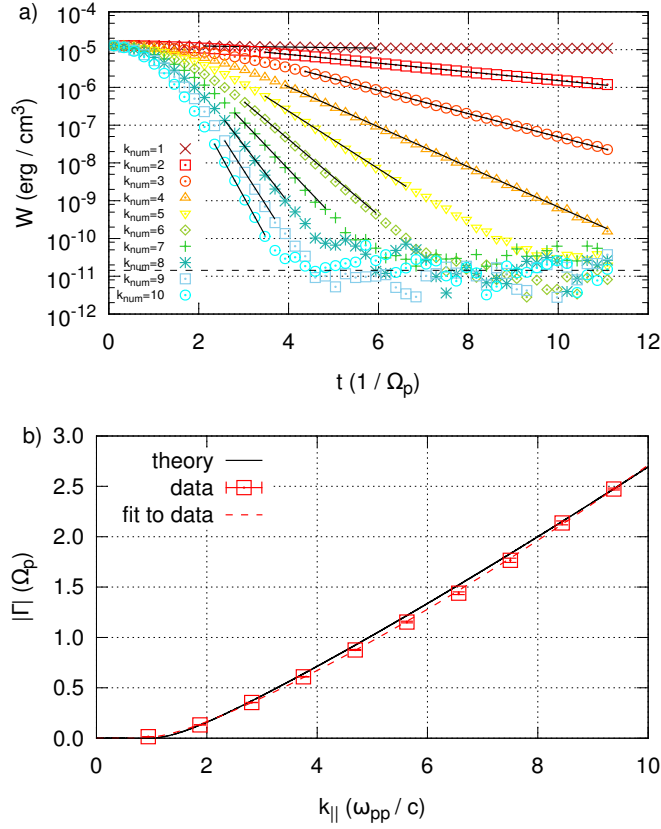}
	\caption{Simulation data and derived damping rates for ten waves with $k_\mathrm{num} = 1 - 10$. Panel a) shows the energy densities $W$ of the waves over time. The dashed line is the average of the energy density at $k_\mathrm{num} = 1 - 10$ in a simulation without wave excitation. Black solid lines are fitted exponential functions which yield the damping rate $\Gamma$. The obtained damping rates (including statistical errors from the fits) are plotted over $k_\parallel$ in panel b). The data can be fitted with the simple fit function from Eq. (\ref{eq:gamma_fit}) and compared to theoretical predictions from Eqs. (\ref{eq:warm_disp}) and (\ref{eq:warm_disp_maxwell}).}
	\label{fig:example_fit}
\end{figure}

Selecting only the data in the exponential phase of wave decay, we then use a least squares exponential fit to obtain the damping rate.
Depending on the data set, each fit for a single wave can be based on a different number of data points from the dispersion relations.
The result of the fitting process is presented in Fig. \ref{fig:example_fit} a), where data and fits for ten waves from two simulations are shown.
In Fig. \ref{fig:example_fit} b) the damping rates $\Gamma(k_\parallel)$ obtained from the fit functions are plotted over the respective $k_\parallel$.

To judge the quality of our measurement we compare our results to the predictions of warm plasma theory.
The theoretical curve $\Gamma_\mathrm{theory}(k_\parallel)$ from Eqs. (\ref{eq:warm_disp}) and (\ref{eq:warm_disp_maxwell}) is shown in Fig. \ref{fig:example_fit} b).
Having used the physical parameters from Table \ref{tab:setup_phys} as input parameters for the PiC simulations and Eqs. (\ref{eq:warm_disp}) and (\ref{eq:warm_disp_maxwell}), the measured rates should be in agreement with $\Gamma_\mathrm{theory}(k_\parallel)$ if cyclotron damping is represented correctly in the simulation.

We further employ a fit function to fit our $\Gamma (k_\parallel)$ \cite{li_2001}:
\begin{equation}
	\frac{\Gamma}{\Omega_\mathrm{p}} = - m_1 \, \left(\frac{k_\parallel \, c}{\omega_\mathrm{p,p}}\right)^{m_2} \cdot \exp{\left(- m_3 \, \frac{\omega_\mathrm{p,p}^2}{k^2 \, c^2}\right)},
	\label{eq:gamma_fit}
\end{equation}
where $m_1$,  $m_2$ and $m_3$ are the fit parameters.
This simplistic function yields a qualitatively and quantitatively accurate approximation of the actual damping rate computed from warm plasma theory.
For example, if fitted to the theoretical values $\Gamma_\mathrm{theory}(k_\parallel)$ at the ten different $k_\parallel$ we chose in the simulations, the full theoretical curve and the fit could hardly be distinguished in Fig. \ref{fig:example_fit} b) -- which is why this fit is not included in the plot.

The more interesting test case is to apply the simple fit function (\ref{eq:gamma_fit}) to our measurement.
Fluctuations in the measured data are averaged out in the fit curve, thus giving a better overall representation of $\Gamma$ over $k_\parallel$.
The newly obtained fit curve can then be compared to theory, or simply be used to interpolate between the points of measured data.

\subsection{Lorentz profiles}
\label{sec:lorentz}

We test the method described above against the approach mentioned in Sect. \ref{sec:introduction}, namely the description of a wave's energy distribution $W(\omega)$ in frequency space by use of a Lorentz profile.
For doing so, we first produce dispersion plots and then take the energy densities $W_{k_\mathrm{num}}(\omega_\mathrm{num})$ for each $k_\mathrm{num}$ representing one of the ten waves in the simulations discussed previously.
We find that the data is strongly influenced by the number of time steps (i.e. the interval) used to produce the dispersion plots and by the points in time at which this interval starts and ends (i.e. early or late times during the simulation).

Looking at Fig. \ref{fig:example_fit} a) and at the positions of the exponential fits therein, we produce two dispersion plots using two intervals of time.
The first (``early'') interval starts at $t_\mathrm{s} \, \Omega_\mathrm{p} = 2.24$ and ends at $t_\mathrm{e} \, \Omega_\mathrm{p} = 5.60$ (15000 time steps).
The second (``late'') one contains the period of time from $t_\mathrm{s} \, \Omega_\mathrm{p} = 4.48$ to $t_\mathrm{e} \, \Omega_\mathrm{p} = 7.84$ (also 15000 time steps).
We use the early interval to obtain the data for $k_\mathrm{num} = \{6, \, 7, \, 8, \, 9, \, 10\}$ and the late interval for $k_\mathrm{num} = \{1, \, 2, \, 3, \, 4, \, 5\}$, which is roughly in accordance with the exponential fits in Fig. \ref{fig:example_fit} a).

\begin{figure}[htb]
	\centering
	\includegraphics[width=0.6\linewidth]{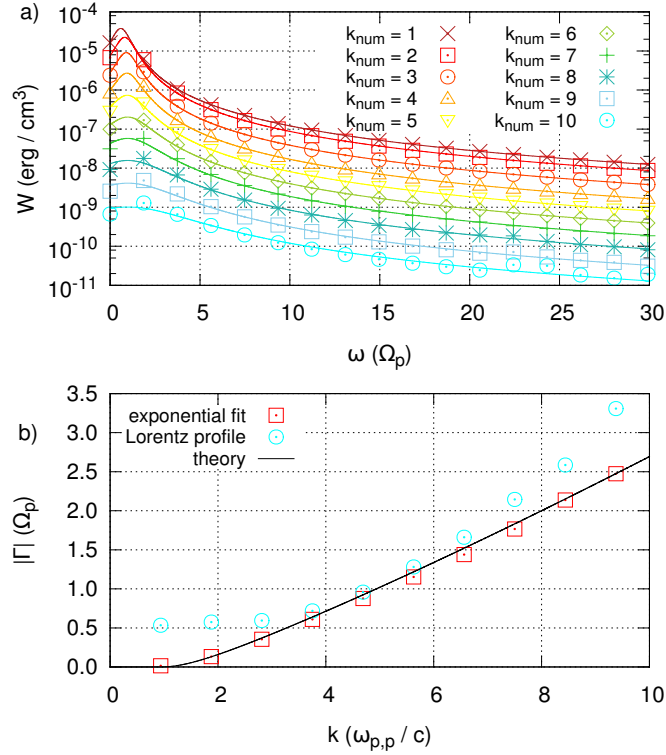}
	\caption{
		Panel a): Energy density $W$ as a function of frequency $\omega$ for ten wave numbers $k_\mathrm{num}$, as indicated.
		The lines represent Lorentz profiles fitted to the data according to Eq. (\ref{eq:lorentz}).
		Panel b): The damping rate $\Gamma$ as a function of the wave number as predicted by theory (solid line) and as measured in the simulation.
		Damping rates are obtained from simulation data using the method described in this article (red squares, labeled ``exponential fit'') and via fitting Lorentz profiles (blue circles, labeled ``Lorentz profile''), as shown in panel a).
 	}
	\label{fig:lorentz}
\end{figure}

A Lorentz profile, Eq. (\ref{eq:lorentz}), can be fitted to the energy density $W_{k_\mathrm{num}}(\omega_\mathrm{num})$ of each wave.
We treat the amplitude $W_0$ and the damping rate $\Gamma$ as free parameters in the fit, but fix the wave's frequency $\omega_0$ at its theoretical value.
The Lorentz profiles resulting from these fits are shown in Fig. \ref{fig:lorentz} a) together with the data from the simulations.

The damping rates obtained from the Lorentz profiles are plotted over the wave number in Fig. \ref{fig:lorentz} b), together with theoretical predictions from Eqs. (\ref{eq:warm_disp}) and (\ref{eq:warm_disp_maxwell}) and the measured damping rates from Fig. \ref{fig:example_fit} b).
Obviously, the damping rates from the Lorentz profiles overestimate the expected $\Gamma$ at small and at large wave numbers.
It is especially interesting that $\Gamma$ is almost constant for $k \, c / \omega_\mathrm{p,p} \leq 3$.
In the intermediate range ($3 < k \, c / \omega_\mathrm{p,p} < 7$) the results from the Lorentz profiles are of similar quality as those from the method discussed in this article.

It has to be noted that the data in Fig. \ref{fig:lorentz} a) resolves the wave's energy profile $W(\omega)$ unexpectedly well.
Due to the initial excitation of L-mode waves their energy density lies well above the noise limit and even drowns the signal of whistler waves, which has to be expected in the frequency regime $0 < \omega < |\Omega_\mathrm{e}|$ and thus in the range plotted in Fig. \ref{fig:lorentz} a).

\section{Effects of the simulation's dimensionality}
\label{sec:dimensions}

After having demonstrated its functional principle we apply our method of obtaining damping rates to simulations with different dimensionality.
For PiC simulations it is common to reduce the simulation box to two or even one spatial dimension(s), if the physical problem permits such a lower-dimensional treatment.
In such a case, particle velocities and electromagnetic fields can still be treated as three-dimensional vectors.
However, the particles may then only move along those directions which are still spatially resolved.

For the study of cyclotron damping of parallel propagating waves only one spatial dimension is essential, namely the direction parallel to the background magnetic field along which the waves propagate.
We therefore set up simulations using the physical and numerical parameters given in Tables \ref{tab:setup_phys} and \ref{tab:setup_num} in Sect. \ref{sec:setup} in one-, two-, and three-dimensional simulation boxes.
As in the example simulations discussed in Sect. \ref{sec:damping_rate} we carry out sets of two simulations each, where waves with even $k_\mathrm{num}$ are excited in one simulation, and waves with odd $k_\mathrm{num}$ in the other.

The specific initial conditions of a simulation are the result a random distribution of the particles and their velocities.
Therefore, by chance, it may be that those random conditions cause the simulation to yield atypical results, which is why one should in principle repeat each simulation with a different random particle distribution.
To reduce the influence of the initial conditions on the final results of the simulations, we repeat each simulation six times, each time with a different seed for the random number generator which produces the initial positions and velocities of the particles.
That means that we perform six (random seeds) times two (even / odd wave numbers) times three (dimensionality) simulations.

To be able to refer to the individual simulations, we define the following nomenclature:
Three-dimensional simulations will be referred to as set A, two-dimensional simulations as set B and one-dimensional simulations as set C.
All simulations of a set which include excited waves with odd (even) $k_\mathrm{num}$ will be denoted by an index 1 (2), e.g. $\mathrm{A}_1$ ($\mathrm{A}_2$).
Within a set, field data from all simulations of the same type (e.g. all simulations of type $\mathrm{A}_1$) is averaged.

In Sect. \ref{sec:dispersion_plots} we have defined the interval $t_\mathrm{int}$ (measured in time steps $\Delta t$) which defines the number of time steps included in the dispersion plots (see Fig. \ref{fig:example_disp}) and the time resolution for the energy density $W(t)$ (see Fig. \ref{fig:example_evolution}).
To test the effect of $t_\mathrm{int}$ we perform the analysis of our simulations with three different interval lengths $t_\mathrm{int} = 1000 \, \Delta t$ ($= 0.22 \, \Omega_\mathrm{p}^{-1}$), $t_\mathrm{int} = 3000 \, \Delta t$ ($= 0.67 \, \Omega_\mathrm{p}^{-1}$), and $t_\mathrm{int} = 5000 \, \Delta t$ ($= 1.12 \, \Omega_\mathrm{p}^{-1}$).

Starting with the three-dimensional simulations of set A, we present the results of our study on the effect of the simulation's dimensionality in the following sections.

\subsection{Three-dimensional simulations}
\label{sec:3D}

\subsubsection{Particle statistics and temperature}
\label{sec:temperature}

In a setup which aims at transferring energy from the electromagnetic fields of plasma waves to the particles, it is especially worthwhile to take a look at the particle population and its velocity spectrum.
Hence, before we start with analyzing the damping rates of plasma waves, we take a look at the velocity distributions of protons and electrons.
At different points in time during the simulation, the full particle data is stored for later examination.
Studying the development of the velocity spectra of both electrons and protons yields information about temperature changes and other kinetic effect during the simulation.

As stated before, we consider a thermal plasma, which means that the spectrum of each velocity component (in Cartesian coordinates $v_{\mathrm{s},x}$, $v_{\mathrm{s},y}$ and $v_{\mathrm{s},z}$, with 's' denoting the particle species) follows a Gaussian distribution and the spectrum of the absolute of the velocities ($v_\mathrm{s} = |\vec{v}_\mathrm{s}|$) follows a Maxwell-Boltzmann distribution.
Protons and electrons can be analyzed separately.
Velocity data is binned, giving the particle number per bin as a function of $v_\mathrm{s}$ or a component of $\vec{v}_\mathrm{s}$, which then can be fitted by the respective distribution function.
This procedure yields a temperature $T_\mathrm{s}$, as well as temperatures $T_{\parallel,\mathrm{s}}$ and $T_{\perp,\mathrm{s}}$ parallel and perpendicular to the background magnetic field $\vec{B}$ for each species.
As for $T_{\parallel,\mathrm{s}}$ and $T_{\perp,\mathrm{s}}$, the former is produced from the one velocity component parallel to $\vec{B_0}$, whereas the latter is the average over the two temperatures obtained from the spectra of the two velocity components perpendicular to $\vec{B_0}$.

\begin{figure}[ht]
	\centering
	\includegraphics[width=0.6\linewidth]{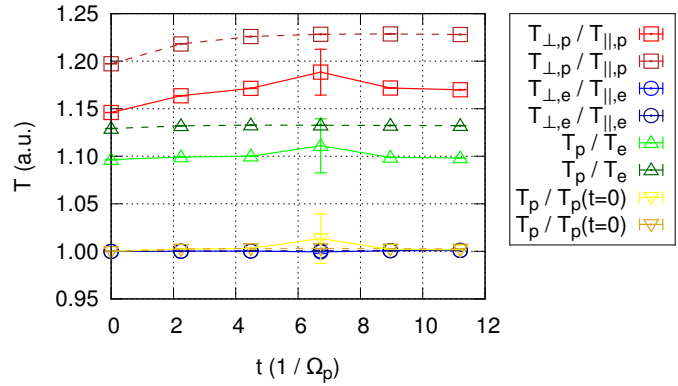}
	\caption{Proton and electron temperature characteristics throughout the simulations. All data is averaged over simulations $\mathrm{AI}_1$ to $\mathrm{AVI}_1$ (bright colors, solid lines) or $\mathrm{AI}_2$ to $\mathrm{AVI}_2$ (darker colors, dashed lines), respectively. Indices 'p' and 'e' refer to protons and electrons, '$\parallel$' and '$\perp$' refer to components parallel and perpendicular to the background magnetic field. See text for details.}
	\label{fig:temperature}
\end{figure}

Figure \ref{fig:temperature} shows temperature development over time.
In this plot, all temperatures are averages over six simulations, where we have analyzed the data of simulations $\mathrm{A}_1$ and $\mathrm{A}_2$ separately.
Errors are calculated from the standard errors of the temperatures derived from the data of every single simulation.

The plot shows that at the beginning of the simulation an anisotropy in the proton temperature is evident, with $T_{\perp,\mathrm{p}}$ being larger than $T_{\parallel,\mathrm{p}}$.
This is a result of the initialization mechanism used for the excitation of waves: particles are initialized with a thermal spectrum, but are then pushed according to the electromagnetic fields of the excited waves.
This leads to larger velocity components perpendicular to $\vec{B_0}$, while the parallel velocity component is not affected.
The effect increases for waves with frequencies closer to the resonance at $\Omega_\mathrm{p}$, which is the reason for a more significant anisotropy in the data from $\mathrm{A}_2$ as compared to $\mathrm{A}_1$.
Electrons are barely affected, though, as can be seen in Fig. \ref{fig:temperature}.

The additional energy given to the proton population during initialization also leads to the proton temperature being higher than the electron temperature.

In the course of the simulation, the anisotropy in proton temperature increases further, which suggests that field energy from the decaying waves is transferred to the protons.
Again, electrons are not affected, as is expected for a proton cyclotron resonance.
With the protons' energy increasing, the temperature $T_\mathrm{p}$ also increases as compared to the temperature at $t \, \Omega_\mathrm{p}=0$.
The electron temperature stays constant throughout the simulation and is therefore not shown in Fig. \ref{fig:temperature}.

Note that the proton temperature and the anisotropy therein rises fastest at the beginning of the simulation ($t \, \Omega_\mathrm{p}=0$ to $t \, \Omega_\mathrm{p} = 6$) and then remains relatively constant.
This suggests that the energy transfer is faster at the beginning of the simulation and slows down later on, which is also expected from an exponential decay of the waves' electromagnetic fields.

Additional information is given in Appendix \ref{app:velocity_spectra}, where we show velocity spectra from one of the simulations of set A.
These spectra illustrate the anisotropy of proton velocities at the beginning and at the end of the simulation and support the results presented in this section.

\subsubsection{Damping rates}
\label{sec:damping_rate_sim}

We investigate the damping rates of the different waves with numerical wave numbers $k_\mathrm{num} = \{1, 2, 3, 4, 5, 6, 7, 8, 9, 10\}$ in set A.
As a reference, the theoretical damping rates are calculated from Eqs. (\ref{eq:warm_disp}) and (\ref{eq:warm_disp_maxwell}), where two sets of parameters are used.
While the plasma frequency and the cyclotron frequencies of protons and electrons can be taken from Table \ref{tab:setup_phys} in Sect. \ref{sec:setup}, the temperature has to be corrected, as Sect. \ref{sec:temperature} has shown that an anisotropy of the proton temperature is evident.
Since the anisotropy is different in simulations $\mathrm{A}_1$ and $\mathrm{A}_2$, two values $(T_{\perp, \mathrm{p}} / T_{\parallel,\mathrm{p}})_1 = 1.15$ and $(T_{\perp, \mathrm{p}} / T_{\parallel,\mathrm{p}})_2 = 1.20$ are chosen according to the data in Fig. \ref{fig:temperature} at $t = 0$.
Similarly, the absolute temperature of the protons has to be corrected to $(T_\mathrm{p} / T_\mathrm{e})_1 = 1.10$ and $(T_\mathrm{p} / T_\mathrm{e})_2 = 1.13$.
The resulting damping rates from Eqs. (\ref{eq:warm_disp}) and (\ref{eq:warm_disp_maxwell}) are still not expected to describe simulation results perfectly, since the temperature changes throughout the simulation, as was shown in Fig. \ref{fig:temperature}.
The theoretical expectations are given in Table \ref{tab:damping_rate_A} in Appendix \ref{app:setA}.

To evaluate the simulation data, dispersion relations are created from the electromagnetic fields of all simulations of set A, as described in Sect. \ref{sec:dispersion_plots}.
The energy density of each excited wave is extracted from the dispersion plots for different points in time in the simulations.
The total energy density $W(k)$ of a wave with wave number $k$ is obtained by summing up the components perpendicular to the background magnetic field $\vec{B_0}$ of both the electric and magnetic fields.
Afterwards a mean energy density $\bar{W}(k)$ is calculated by averaging over the energy densities obtained from the individual simulations of set A.
Note that set A has to be subdivided into $\mathrm{A}_1$ and $\mathrm{A}_2$ since waves with odd $k_\mathrm{num}$ are excited only in set $\mathrm{A}_1$ and waves with even $k_\mathrm{num}$ are only present in set $\mathrm{A}_2$.

In a similar manner an average of the energy density $\bar{W}_\mathrm{noise}(k)$ of the background noise can be obtained.
Since waves with even (odd) $k_\mathrm{num}$ are not excited in simulations $\mathrm{A}_1$ ($\mathrm{A}_2$), the background energy density at each relevant $k$ can be extracted from the dispersion relations as well.
A corrected energy density of the excited waves is then obtained simply by subtracting $\bar{W}_\mathrm{noise}(k)$ from $\bar{W}(k)$.
For the sake of less confusing notation, the energy density will be denoted by $W(k)$ throughout the article, although the corrected and averaged energy density is used when simulation data is presented.

\begin{figure}[htb]
	\centering
	\includegraphics[width=0.6\linewidth]{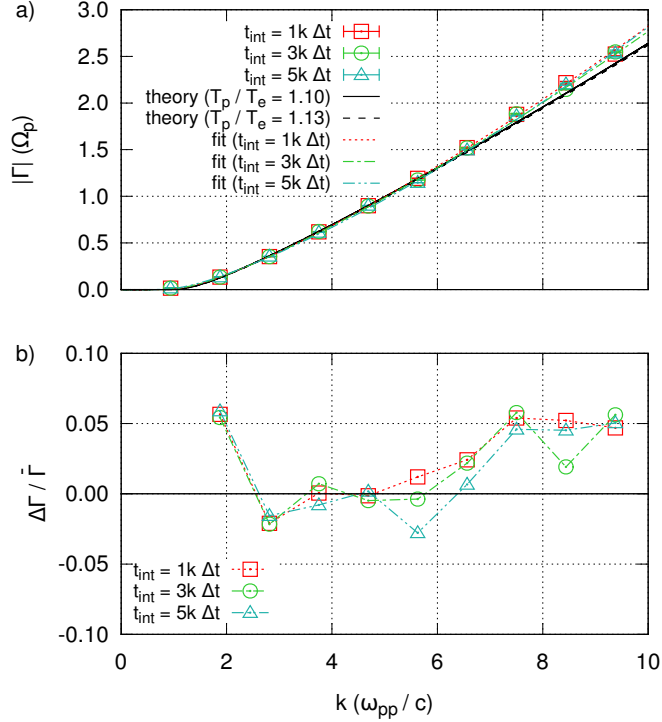}
	\caption{
		Measured and theoretical damping rates $\Gamma$ (panel a) and relative difference thereof (panel b) as functions of the parallel wave number $k$.
		Panel a): Fits to the simulation data are performed according to Eq. (\ref{eq:gamma_fit}); error bars represent standard errors.
		Panel b): $\Delta\Gamma / \bar{\Gamma}$ is calculated according to Eq. (\ref{eq:gamma_deviation}); data at $k_\mathrm{num} = 0$ is excluded due to the large deviation from theory (see text).
	}
	\label{fig:damping_rate_A}
\end{figure}

Having obtained the energy density $W(k)$ of each of the excited waves in the simulations of set A, the damping rates can be computed using the method described in Sect. \ref{sec:energy_density}.
These rates are also given in Table \ref{tab:damping_rate_A} in Appendix \ref{app:setA}.
A plot representing the data is presented in Fig. \ref{fig:damping_rate_A} a).
Three different interval lengths $t_\mathrm{int}$ have been used for data evaluation, meaning that we have used dispersion relations built from $t_\mathrm{int} = \{1000, \, 3000, \, 5000\}$ time steps $\Delta t$ to extract the energy densities of the electromagnetic fields of the excited waves.
Plots showing the energy densities $W$ over time for all ten waves and the three interval lengths can be found in Appendix \ref{app:setA}.

Looking at the data given in Table \ref{tab:damping_rate_A} in Appendix \ref{app:setA}, it can be seen that the deviation between the theoretical predictions for the two temperature settings is of the order of less than one percent (except for $k_\mathrm{num} = 1$).
Furthermore, it becomes obvious that the errors of the measured damping rates are typically much smaller than the actual deviation of measured data and theoretical prediction.
The errors given are computed from the standard errors of the averaged electromagnetic fields obtained from the dispersion plots via Gaussian error propagation.
However, this method seems not to capture the actual deviation from theory.
We therefore calculate the deviation
\begin{equation}
	\Delta\Gamma / \bar{\Gamma} = \frac{\Gamma_\mathrm{sim} - \Gamma_\mathrm{theory}}{0.5 \, \big(\Gamma_\mathrm{sim} + \Gamma_\mathrm{theory}\big)}
	\label{eq:gamma_deviation}
\end{equation}
of measured and theoretical damping rate and present the results in Fig. \ref{fig:damping_rate_A} b).
Note that different theoretical predictions are used for odd and even $k_\mathrm{num}$, as described above.

Both the curves in Fig. \ref{fig:damping_rate_A} a) and Fig. \ref{fig:damping_rate_A} b), showing the damping rate $\Gamma$ and its deviation from theory as a function of the wave number $k$, suggest that our measurements are in good agreement with theoretical prediction over a wide range of wave numbers.
Judging only from observations with the naked eye, all measured damping rates in Fig. \ref{fig:damping_rate_A} are in perfect agreement with theory up to $k \, c / \omega_\mathrm{pp} \sim \! 7$ -- a claim which is also supported by the fits according to Eq. \ref{eq:gamma_fit}.
At larger $k$ the measured damping rates lie above the theoretical predictions.
Note that the fits are performed using all data, neglecting the different temperature settings in simulations $\mathrm{A}_1$ and $\mathrm{A}_2$.
Fit parameters are given in Table \ref{tab:gamma_fit_A}.

\begin{table}[htb]
	\centering
	\begin{tabular}{c c c c}
		\hline
		\noalign{\smallskip}
		$t_\mathrm{int} \, (\Delta t)$ & $m_1$ & $m_2$ & $m_3$ \\
		\noalign{\smallskip}
		\hline
		\noalign{\smallskip}
		1000 & $0.118 \pm 0.004$ & $1.391 \pm 0.017$ & $2.64 \pm 0.11$ \\
		\noalign{\smallskip}
		3000 & $0.119 \pm 0.008$ & $1.38 \pm 0.03$ & $2.6 \pm 0.3$ \\
		\noalign{\smallskip}
		5000 & $0.101 \pm 0.012$ & $1.45 \pm 0.06$ & $2.1 \pm 0.5$ \\
		\noalign{\smallskip}
		\hline
	\end{tabular}
	\caption{Fit parameters including standard errors for the fits to the data of set A in Fig. \ref{fig:damping_rate_A} a) according to Eq. (\ref{eq:gamma_fit}).}
	\label{tab:gamma_fit_A}
\end{table}

Taking a closer look at the relative deviations in Fig. \ref{fig:damping_rate_A} b), it can be confirmed that the simulations yield an accurate representation of cyclotron damping.
In between $\sim \! 2 \, < \, k \, c / \omega_\mathrm{pp} \, < \, \sim \! 7$ the deviation $\Delta \Gamma / \bar{\Gamma}$ is mainly below two percent.
This is especially interesting to notice, since the deviation between theoretical predictions for different proton temperatures is in the order just short of one percent.
With the plasma being heated during the simulations -- which makes theoretical predictions only an estimate -- it can be expected that measured data and theoretical prediction will differ at least by the same amount as theoretical predictions for the temperatures at the beginning and the end of the simulation.
Keeping that in mind, a deviation below two percent appears to be perfectly reasonable and accurate.
The deviations of simulation data and theory for $k \, c / \omega_\mathrm{pp} \, < \, 2$ and $k \, c / \omega_\mathrm{pp} \, > \, 7$ is around five percent, which still yields an appropriate estimate of the damping rate in these regimes.
Note that the data point at $k_\mathrm{num} = 1$ ($k \, c / \omega_\mathrm{pp} \sim 0.9$) is not shown in Fig. \ref{fig:damping_rate_A} b).
This case will be treated separately in Appendix \ref{app:weak_damping}, since the time evolution of the energy density of the wave with $k_\mathrm{num} = 1$ hints at unforeseen effects which hinder the measurement of the damping rate.
For the following studies of one- and two-dimensional simulations in Sects. \ref{sec:1D} and \ref{sec:2D} the wave with $k_\mathrm{num} = 1$ will also be excluded.

The results presented in Fig. \ref{fig:damping_rate_A} suggest that the length of the interval $t_\mathrm{int}$ do not influence the overall results.
The idea behind the different interval lengths is to trade temporal resolution (small $t_\mathrm{int}$) for more meaningful averages of the energy density over time (large $t_\mathrm{int}$), suppressing fluctuations on short time scales.
However, the results are similar in all cases and thus no approach is preferred above the other two.

\subsection{Two-dimensional simulations}
\label{sec:2D}

In this section we present data obtained from set B, a set of twelve two-dimensional simulations analogous to set A (see Sect. \ref{sec:3D}).
Independent of the dimensionality of the simulation, the method to determine the damping rates of excited waves is still the same.

\begin{figure}[htb]
	\centering
	\includegraphics[width=0.6\linewidth]{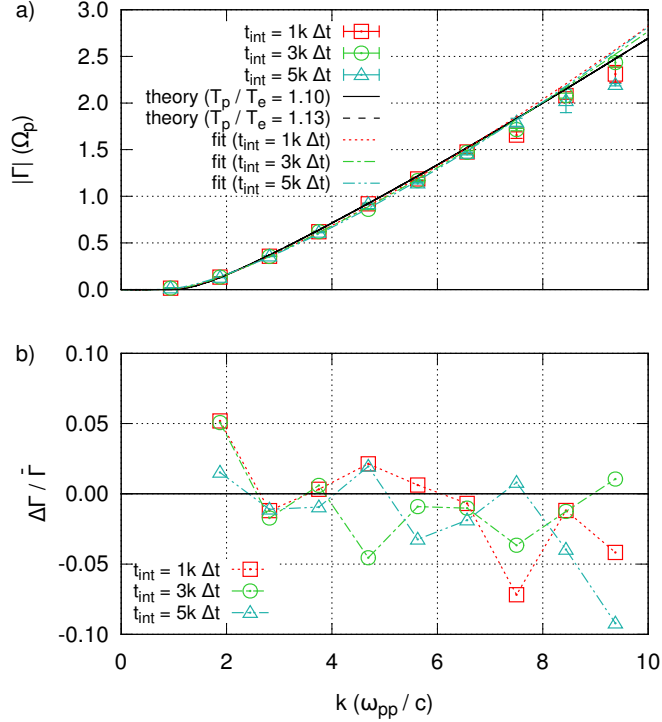}
	\caption{
		Measured and theoretical damping rates $\Gamma$ (panel a) and relative difference thereof (panel b) as functions of the parallel wave number $k$ as obtained from simulations B.
		Data at $k_\mathrm{num} = 1$ is omitted in this plot.
		Panel a): Fits to the simulation data are performed according to Eq. (\ref{eq:gamma_fit}); error bars represent standard errors.
		Panel b): $\Delta\Gamma / \bar{\Gamma}$ is calculated according to Eq. (\ref{eq:gamma_deviation}).
	}
	\label{fig:damping_rate_B}
\end{figure}

The obtained damping rates are shown in Table \ref{tab:damping_rate_B} in Appendix \ref{app:setB}.
Like set A, set B consists of two sub-sets $\mathrm{B}_1$ and $\mathrm{B}_2$ in which waves with odd and even $k_\mathrm{num}$ are excited.
The damping rates $\Gamma$ are plotted over the wave number $k$ in Fig. \ref{fig:damping_rate_B} a), together with fits according to Eq. (\ref{eq:gamma_fit}).
Fit parameters are given in Table \ref{tab:gamma_fit_B}.
Figure \ref{fig:damping_rate_B} b) shows the relative deviation $\Delta \Gamma / \bar{\Gamma}$ of measured data and theoretical expectations according to Eq. (\ref{eq:gamma_deviation}).

The measured data tend to underestimate the damping rates at high $k$, as Fig. \ref{fig:damping_rate_B} a) suggests.
This is confirmed by Fig. \ref{fig:damping_rate_B} b), which shows a trend towards negative deviation $\Delta \Gamma / \bar{\Gamma}$ at $k \, c / \omega_\mathrm{pp} > 6$.
This trend is persistent for all three different interval lengths $t_\mathrm{int}$.
However, in the range $\sim \! 2 \, < \, k \, c / \omega_\mathrm{pp} \, < \, \sim \! 7$ the relative deviation is again mainly below two percent, indicating good agreement of simulation data and theoretical expectations.

The standard errors given in Table \ref{tab:damping_rate_B} in Appendix \ref{app:setB} approach the actual deviation from the theoretical predictions, as they are of the order of a few percent over the whole range of wave numbers.

\begin{table}[htb]
	\centering
	\begin{tabular}{c c c c}
		\hline
		\noalign{\smallskip}
		$t_\mathrm{int} (\Delta t)$ & $m_1$ & $m_2$ & $m_3$ \\
		\noalign{\smallskip}
		\hline
		\noalign{\smallskip}
		1000 & $0.155 \pm 0.014$ & $1.24 \pm 0.05$ & $3.3 \pm 0.3$ \\
		\noalign{\smallskip}
		3000 & $0.125 \pm 0.009$ & $1.33 \pm 0.03$ & $2.7 \pm 0.3$ \\
		\noalign{\smallskip}
		5000 & $0.192 \pm 0.016$ & $1.11 \pm 0.04$ & $4.0 \pm 0.4$ \\
		\noalign{\smallskip}
		\hline
	\end{tabular}
	\caption{Fit parameters including standard errors for the fits to the data of set B in Fig. \ref{fig:damping_rate_B} a) according to Eq. (\ref{eq:gamma_fit}).}
	\label{tab:gamma_fit_B}
\end{table}

Measured energy densities $W$ are plotted over time in Fig. \ref{fig:energy_B} for $t_\mathrm{int} = \{1000, \, 3000, \, 5000\} \, \Delta t$.
Here, the impact of the two-dimensional setup becomes obvious in the energy level of background noise $W_\mathrm{noise}$, which is about two orders of magnitude above the noise levels found in set A (see Fig. \ref{fig:energy_A} in Appendix \ref{app:setA}).
This leaves less space for the exponential fits (black lines) and thus potentially reduces the quality of the results.

\subsection{One-dimensional simulations}
\label{sec:1D}

\begin{figure}[htb]
	\centering
	\includegraphics[width=0.6\linewidth]{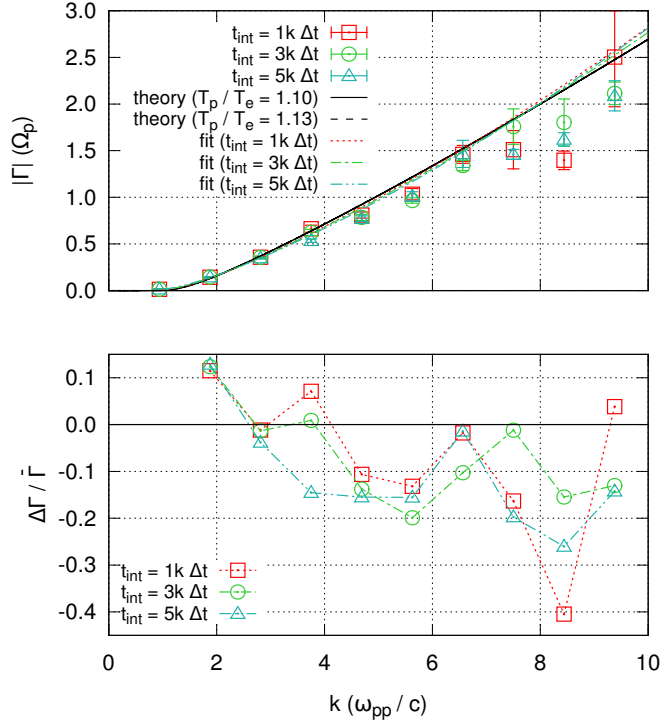}
	\caption{
		Measured and theoretical damping rates $\Gamma$ (panel a) and relative difference thereof (panel b) as functions of the parallel wave number $k$ as obtained from simulations C.
		Data at $k_\mathrm{num} = 1$ is omitted in this plot.
		Panel a): Fits to the simulation data are obtained according to Eq. (\ref{eq:gamma_fit}); error bars represent standard errors.
		Panel b): $\Delta\Gamma / \bar{\Gamma}$ is calculated according to Eq. (\ref{eq:gamma_deviation}).
	}
	\label{fig:damping_rate_C}
\end{figure}

In this section we present data obtained from one-dimensional simulations of set C.
The measured damping rates are shown in Table \ref{tab:damping_rate_C} in Appendix \ref{app:setC}.
As described at the beginning of Sect. \ref{sec:dimensions}, set C consists of two sub-sets $\mathrm{C}_1$ and $\mathrm{C}_2$ in which waves with odd and even $k_\mathrm{num}$ are excited.
The damping rates $\Gamma$ are plotted over the wave number $k$ in Fig. \ref{fig:damping_rate_C} a), together with fits according to Eq. (\ref{eq:gamma_fit}).
Fit parameters are given in Table \ref{tab:gamma_fit_C}.
Figure \ref{fig:damping_rate_C} b) shows the relative deviation $\Delta \Gamma / \bar{\Gamma}$ between measured data and theoretical expectations according to Eq. (\ref{eq:gamma_deviation}).

The measured data tend to underestimate the damping rates even more than in the case of the two-dimensional setup presented in Sect. \ref{sec:2D}.
Figure \ref{fig:damping_rate_C} a) shows that the measured data deviate from the theoretical expectations already at $k\, c / \omega_\mathrm{pp} > 4$ in a systematic manner.
This behavior becomes even more evident when looking at the fits to the data or at the relative deviation plotted in Fig. \ref{fig:damping_rate_C} b).

Contrary to the results from the three- and two-dimensional simulations (see Sects. \ref{sec:3D} and \ref{sec:2D}), the data in Fig. \ref{fig:damping_rate_C} b) do not exhibit a region in $k$ where the relative deviation is at a relatively constant and low level.
Starting already at smallest $k$ a steady downward trend can be observed, leading to deviations of up to $\sim \! 40$ percent.

The standard errors given in Table \ref{tab:damping_rate_C} in Appendix \ref{app:setC} are in the order of a few to more than ten percent over the whole range of wave numbers.
However, in many cases this rather large uncertainty does not cover the theoretical predictions.
Note that the fits to the data also become more and more unreliable, as the errors in Table \ref{tab:gamma_fit_C} show.
Especially parameter $m_3$, which is part of the argument of the exponential function in Eq. (\ref{eq:gamma_fit}), shows standard errors of up to $\sim \! 60$ percent.

\begin{table}[htb]
	\centering
	\begin{tabular}{c c c c}
		\hline
		\noalign{\smallskip}
		$t_\mathrm{int} (\Delta t)$ & $m_1$ & $m_2$ & $m_3$ \\
		\noalign{\smallskip}
		\hline
		\noalign{\smallskip}
		1000 & $0.25 \pm 0.12$ & $0.9 \pm 0.2$ & $4.1 \pm 1.5$ \\
		\noalign{\smallskip}
		3000 & $0.10 \pm 0.02$ & $1.37 \pm 0.11$ & $1.7 \pm 1.0$ \\
		\noalign{\smallskip}
		5000 & $0.13 \pm 0.04$ & $1.23 \pm 0.13$ & $2.3 \pm 1.2$ \\
		\noalign{\smallskip}
		\hline
	\end{tabular}
	\caption{Fit parameters including standard errors for the fits to the data of set C in Fig. \ref{fig:damping_rate_C} a) according to Eq. (\ref{eq:gamma_fit}).}
	\label{tab:gamma_fit_C}
\end{table}

Measured energy densities $W$ are plotted over time in Fig. \ref{fig:energy_C} in Appendix \ref{app:setC} for $t_\mathrm{int} = \{1000, \, 3000, \, 5000\} \, \Delta t$.
Here, the impact of the one-dimensional setup becomes obvious in the energy level of background noise $W_\mathrm{noise}$, which is about four orders of magnitude above the noise levels found in set A (see Fig. \ref{fig:energy_A} in Appendix \ref{app:setA}).
The energy range in which the exponential fits can be applied is thus narrowed down to one to two orders of magnitude in $W$, which reduces the quality of the fits drastically.
It can be expected that the correct slope of the exponential function is not represented in the data, since the region between the slow onset of wave damping and the background noise level is not sufficiently broad.

\section{Comparison of simulations with different parameter sets}
\label{sec:parameters}

In this section, the physical parameters given in Table \ref{tab:setup_phys} in Sect. \ref{sec:setup} are varied, whereas most of the numerical parameters from Table \ref{tab:setup_num} are kept constant.
The resulting set of simulations, set D, consists of six individual simulations ($\mathrm{D}_1$ through $\mathrm{D}_6$), each containing five excited waves.
The change of the background magnetic field $B_0$, the thermal speed $v_\mathrm{th}$, and the plasma frequency $\omega_\mathrm{p}$ leads to a change in the plasma beta (e.g. \cite{huba_2007})
\begin{equation}
	\beta = 8 \, \pi \, n \, k_\mathrm{B} \, T \, / \, B_0^2,
	\label{eq:beta1}
\end{equation}
where $n = n_\mathrm{e} + n_\mathrm{p}$ is the particle number density (electrons and protons) and $k_\mathrm{B}$ is the Boltzmann constant.
Assuming that the plasma temperature is the same for electrons and protons, $T_\mathrm{e} = T_\mathrm{p} = T$, and that $n_\mathrm{e} = n_\mathrm{p}$, the above equation can be written as
\begin{equation}
	\beta = \frac{4 \, \omega_\mathrm{p,e}^2 \, v_\mathrm{th,e}^2}{\Omega_\mathrm{e}^2 \, c^2}.
	\label{eq:beta2}
\end{equation}
We will use this definition of $\beta$ to better describe and refer to the individual simulations of set D.
A list of parameters used for each of the six simulations is presented in Table \ref{tab:setup_D}.
Parameters not listed, such as $N_\parallel$, $N_\perp$ and $m_\mathrm{p} / m_\mathrm{e}$ are the same as in Table \ref{tab:setup_num} in Sect. \ref{sec:setup}.
The number of time steps has been set to $N_t = 50000$ in all simulations except for $\mathrm{D}_3$, where $N_t = 100000$.

\begin{table}[htb]
	\centering
	\begin{tabular}{c c c c c c c}
		\hline
		\noalign{\smallskip}
		simulation & $\omega_\mathrm{p,e} \, (\mathrm{rad} \, \mathrm{s}^{-1})$ & $|\Omega_\mathrm{e}| \, (\omega_\mathrm{p,e})$ & $v_\mathrm{th,e} \, (c)$ & $\beta$ & $\Delta x \, (c \, \omega_\mathrm{p,e}^{-1})$ & $\Delta t \, (\omega_\mathrm{p,e}^{-1})$\\
		\noalign{\smallskip}
		\hline
		\noalign{\smallskip}
		$\mathrm{D}_1$ & $2.0 \cdot 10^8$ & $2.64 \cdot 10^{-1}$ & $0.05$ & $0.144$ & $3.53 \cdot 10^{-2}$ & $2.04 \cdot 10^{-2}$\\
		\noalign{\smallskip}
		$\mathrm{D}_2$ & $2.0 \cdot 10^8$ & $8.79 \cdot 10^{-1}$ & $0.05$ & $0.013$ & $3.53 \cdot 10^{-2}$ & $2.04 \cdot 10^{-2}$\\
		\noalign{\smallskip}
		$\mathrm{D}_3$ & $2.0 \cdot 10^8$ & $4.40 \cdot 10^{-1}$ & $0.02$ & $0.008$ & $1.41 \cdot 10^{-2}$ & $8.16 \cdot 10^{-3}$\\
		\noalign{\smallskip}
		$\mathrm{D}_4$ & $2.0 \cdot 10^8$ & $4.40 \cdot 10^{-1}$ & $0.10$ & $0.207$ & $7.04 \cdot 10^{-2}$ & $4.06 \cdot 10^{-2}$\\
		\noalign{\smallskip}
		$\mathrm{D}_5$ & $5.0 \cdot 10^7$ & $1.76$ & $0.05$ & $0.003$ & $3.53 \cdot 10^{-2}$ & $2.04 \cdot 10^{-2}$\\
		\noalign{\smallskip}
		$\mathrm{D}_6$ & $5.0 \cdot 10^8$ & $1.76 \cdot 10^{-1}$ & $0.05$ & $0.323$ & $3.53 \cdot 10^{-2}$ & $2.04 \cdot 10^{-2}$\\
		\noalign{\smallskip}
		\hline
	\end{tabular}
	\caption{Physical and numerical parameters for the simulations of series D.}
	\label{tab:setup_D}
\end{table}

\begin{figure}[htb]
	\centering
	\includegraphics[width=0.6\linewidth]{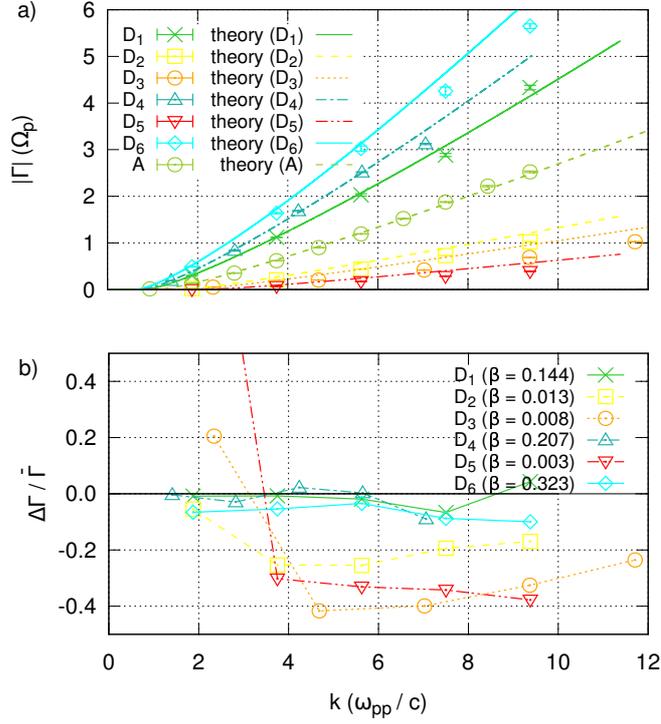}
	\caption{
		Panel a): Measured damping rates $\Gamma$ from simulations with different physical setups (sets A and D, see text for details).
		Theoretical curves are given for comparison.
		Panel b): Relative deviation of measured data and theoretical predictions for set D.
		The first point of data for simulation $\mathrm{D}_5$ lies at $\Delta\Gamma / \bar{\Gamma} = 1.63$, but is cut off in this plot.
	}
	\label{fig:damping_rate_D}
\end{figure}

The measured damping rates for the six simulations of set D are shown together with those obtained from the data of set A in Fig. \ref{fig:damping_rate_D}.
An interval length $t_\mathrm{int} = 1000 \, \Delta t$ has been used.
Lines represent theoretical predictions from Eqs. (\ref{eq:warm_disp}) and (\ref{eq:warm_disp_maxwell}), where $T_\mathrm{e} = T_\mathrm{p} = T$ is assumed.
Fits to the data are omitted here.

Because of the different sets of physical parameters and the resulting $\beta$ in each of the simulations, the dispersion curves and especially the shape of $\Gamma (k)$ also change, as can be seen in Fig. \ref{fig:damping_rate_D} a).
The plot reveals two trends:
First, above $|\Gamma| / \Omega_\mathrm{p} \simeq 3$ the deviation of measured data and theoretical model increases.
This is due to the fast damping of the wave, which makes reliable measurements hard to obtain.
Secondly, the measurements suggest that the simulations underestimate the damping rates at low $\beta$, as can be seen when looking at $\mathrm{D}_2$, $\mathrm{D}_3$ and $\mathrm{D}_5$.
The reason for this second trend is unclear to us, but will be examined in more detail in the following paragraphs.

Figure \ref{fig:damping_rate_D} b) depicts the relative deviation of measured data and theory, with the deviation $\Delta\Gamma / \bar{\Gamma}$ being defined by Eq. (\ref{eq:gamma_deviation}).
This plot makes the disagreement of measurement and theoretical model in the case of low $\beta$ plasmas even more clear.
With the exception of the first point of data, the curves for $\mathrm{D}_2$, $\mathrm{D}_3$ and $\mathrm{D}_5$ suggest an underestimation of the damping rate by 20 to 40 percent.
The first data point for $\mathrm{D}_3$ and $\mathrm{D}_5$ suggests that at very low damping rates, the measurement is not reliable and significantly overestimates the actual damping rate, as discussed at the end of Sect. \ref{sec:damping_rate_sim} and in Appendix \ref{app:weak_damping}.
The simulations with higher $\beta$ exhibit deviations of less than ten percent over the whole range of wave numbers.

\begin{figure}[htb]
	\centering
	\includegraphics[width=0.6\linewidth]{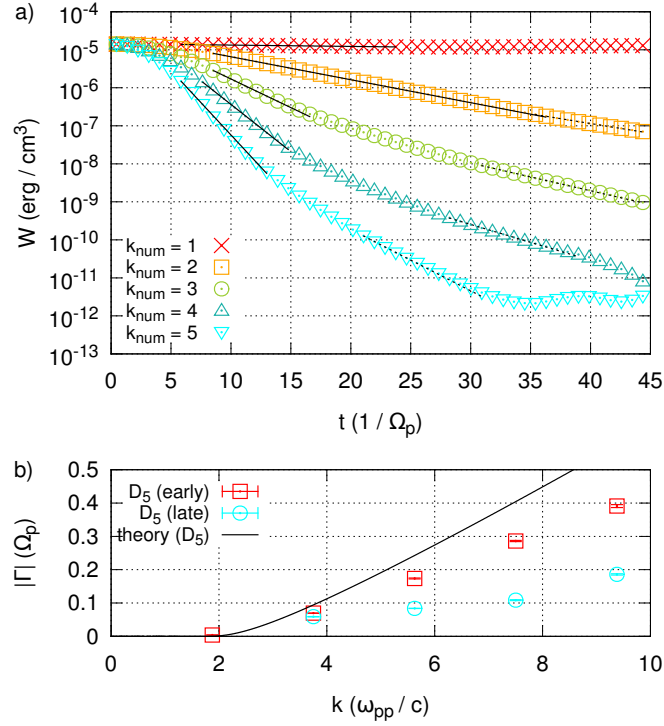}\\
	\caption{
		Panel a): Energy density $W$ of the excited waves in simulation $\mathrm{D}_5$ as a function of time. For most waves two intervals can be found in which an exponential fit can be applied to the data (``early'' and ``late''), as shown by the solid and dotted black lines.
		Panel b): Measured damping rates as a function of the wave number $k$, together with theoretical predictions.
	}
	\label{fig:damping_rate_D5}
\end{figure}

In the case of simulations with a low $\beta$, wave damping is underestimated, as can be seen in Figs. \ref{fig:damping_rate_D} a) and b).
However, the damping rates presented above have been obtained with a very benevolent choice of original data for the exponential fits to the energy density $W$.
In fact, in some cases it would be possible to find a different exponential fit to $W$ later in the simulations, as depicted for simulation $\mathrm{D}_5$ in the top panel of Fig. \ref{fig:damping_rate_D5}.
Here, the exponential fits used above are shown as solid black lines, whereas alternative fits are represented by dotted black lines.
As can be seen, the energy density exhibits the usual transit to an exponential decay at the beginning of the simulation, as discussed in Sect. \ref{sec:damping_rate}.
At later times in the simulation, the exponential decay slows down and eventually transitions into a region with a different slope which can be interpreted as exponential decay with a different damping rate.

Panel b) of Fig. \ref{fig:damping_rate_D5} shows the measured damping rates together with theoretical predictions.
The terms ``early'' and ``late'' refer to the intervals in simulation time, in which the damping rates have been determined, i.e. the solid and dotted lines in the top panel of Fig. \ref{fig:damping_rate_D5}.
It can be easily seen that the agreement of measurement and model is even worse when the ``late'' damping rates are considered.
Unfortunately, the reason for the behavior of the waves' energy densities throughout the simulation are unknown to us.

\section{Discussion and conclusions}
\label{sec:discussion}

In this article we have suggested a simple test setup for simulating cyclotron damping using a PiC approach and presented a method to obtain the damping rate of a given wave from simulation data, equivalent to the method of Koen et al. \cite{koen_2012}.
As long as electromagnetic field data is produced and saved to disk several times during the simulation, the method described in Sect. \ref{sec:damping_rate} can be used to derive the damping rate of a wave from dispersion plots produced from the field data at different points in time.
If Fourier transformed ($k$-space) field data is available at run-time, it is even sufficient to write only the field data (or energy) at specific wave numbers $k$, that are of interest for later analysis, to disk and thus reduce the required output dramatically.
Our method is not necessarily limited to PiC simulations, but can in principle be used with any numerical approach to plasma physics -- as long as the approach used supports the interactions leading to cyclotron damping, of course.

Using our mechanism for wave excitation (see Sect. \ref{sec:setup} or \cite{schreiner_2014_a}) we set up two example simulations which are then evaluated in Sect. \ref{sec:damping_rate} in order to give some practical insight into the analysis needed to obtain damping rates.
Our analysis shows that the excitation of several left-handed, circularly polarized, low frequency waves (in the frequency range between Alfv\'en and ion cyclotron waves) leads to cyclotron damping by thermal protons, as expected.
Cyclotron damping leads to a transfer of energy from the electromagnetic fields of the damped waves to the resonant particles.
Thus, a characteristic decline of the energy density $W$ of the wave can be found.
With the method described in Sect. \ref{sec:damping_rate}, we are able to reproduce the energy density $W(t)$ of each excited wave as a function of time.
The application of an exponential fit to $W(t)$ then yields the damping rate $\Gamma$.

Repeating this procedure for several waves at different wave numbers $k_\parallel$ yields an estimate for $\Gamma(k_\parallel)$, which can be compared to theoretical predictions from warm plasma theory.
Results from our example simulations exhibit an acceptable overall agreement with theory, as Fig. \ref{fig:example_fit} b) in Sect. \ref{sec:energy_density} shows.
This result is particularly interesting, since the excited waves on the low frequency branch of the L-mode are not resolved in the dispersion plots.
Finding the correct damping rate can therefore be seen as a proof for the correct behavior of these waves in the simulation.

For comparison, we determine the the damping rates of the same waves again, using a different method.
Using Lorentz profiles to describe the spectral energy distribution $W(\omega)$ of a wave yields the damping rate, which is the half width at half maximum of the distribution.
The results obtained in Sect. \ref{sec:lorentz} suggest that this method is less reliable than the new method characterized in this article.

Better results from the Lorentz profiles are expected if a higher spectral resolution of the waves in question is available and both flanks of the distribution can be seen.
However, for the case of damped L-mode waves at low frequencies, such a high spectral resolution cannot be achieved, since resolution increases with the run time of the simulation.
Run time cannot be increased indefinitely, as the waves will be dissipated completely after a finite amount of time.

The method of Koen et al. \cite{koen_2012} does not suffer from low spectral resolution, as shown in Sect. \ref{sec:energy_density}.
As has been demonstrated in our example, damping rates can even be obtained if the frequency of the respective wave is not resolved at all.
This is, of course, limited to the case in which only one wave mode is under-resolved.
As soon as the spectral resolution of the dispersion plots is so low that several wave modes are mapped into the same (or the lowest) frequency bin, the wave modes can no longer be distinguished and our method to determine damping rates breaks down.

Of course, our method can also be applied to right-handed waves on the whistler branch or electron cyclotron waves.
In this case, it should be possible to extract the frequency and the damping rate and thus compare both to the theoretical dispersion relation.
So far, we have only studied purely parallel propagating waves.
However, it should also be possible to use the method described in Sect. \ref{sec:damping_rate} to determine damping rates of oblique waves, although obtaining accurate theoretical predictions might be more complicated in this case.

In Sect. \ref{sec:dimensions} we wave tested the influence of the simulation's spatial dimensions on the representation of waves and cyclotron damping.
Since PiC simulations require relatively large amounts of computing time, it is often attempted to reduce the computational cost by reducing the spatial dimensions in the simulation.
In theory, this can also be done in the case of cyclotron damping of parallel propagating waves, since only the direction parallel to the background magnetic field $B_0$ has to be resolved.
Thus, two- or even one-dimensional simulations are possible, as long as electromagnetic fields and particle velocities are still treated as three-dimensional vectors.

The comparison of results from three- (Sect. \ref{sec:3D}), two- (Sect. \ref{sec:2D}) and one-dimensional (Sect. \ref{sec:1D}) simulations shows that the accuracy of the measurement of the damping rate decreases drastically when the dimensionality of the simulation is reduced.
The key problem is the reduced number of particles in simulations with less spatial dimensions, which leads to an increase in background noise and thus to a reduced signal to noise ratio.

We have kept the number of particles per cell constant while decreasing the number of cells in the simulation, thus reducing the total particle count.
One could also increase the number of particles per cell to maintain the total number of particles, i.e. perform a one- or two-dimensional simulation with the same number of particles as a corresponding three-dimensional simulation.
In this latter case we expect better results from two- or one-dimensional simulations.
However, such simulations are often not feasible, since the computational effort -- to first order -- scales with the total number of particles per simulation.
Thus, the advantage of a two-dimensional simulation vanishes, if the particle count is as high as it would be in three dimensions.

Lastly, in Sect. \ref{sec:parameters} we present results from three-dimensional simulations with different sets of physical parameters.
We find that our simulations reproduce the expected cyclotron damping with reasonable deviation from theory only in a certain range of plasma betas.
As Fig. \ref{fig:damping_rate_D} shows, the deviations from theory increase for small plasma betas.

Overall, we argue that PiC simulations are capable of reproducing cyclotron damping correctly.
However, not every physical or numerical configuration might be suitable for this endeavor.
The article at hand concentrates more on the general method of determining the damping rate from a set of simulation data and compares results of a few different simulations.
However, it might also be worthwhile to investigate wave damping with different numerical schemes, such as hybrid PiC / MHD codes or Vlasov codes, and compare these numerical approaches.

\section*{Acknowledgments}
The authors gratefully acknowledge the Gauss Centre for Supercomputing e.V. (www.gauss-centre.eu) for funding this project by providing computing time on the GCS Supercomputer SuperMUC at Leibniz Supercomputing Centre (LRZ, www.lrz.de).

This work is based upon research supported by the National Research Foundation and Department of Science and Technology.
Any opinion, findings and conclusions or recommendations expressed in this material are those of the authors and therefore the NRF and DST do not accept any liability in regard thereto.

We acknowledge the use of the \emph{ACRONYM} code and would like to thank the developers (Verein zur F\"orderung kinetischer Plasmasimulationen e.V.) for their support.

The authors would like to thank Andreas Kempf for providing computational resources for some of the early test simulations and for many helpful comments on and optimization of the piece of code which computes the warm plasma dispersion relations.

\appendix

\section{Velocity spectra}
\label{app:velocity_spectra}

\begin{figure}[ht]
	\centering
	\includegraphics[width=0.8\linewidth]{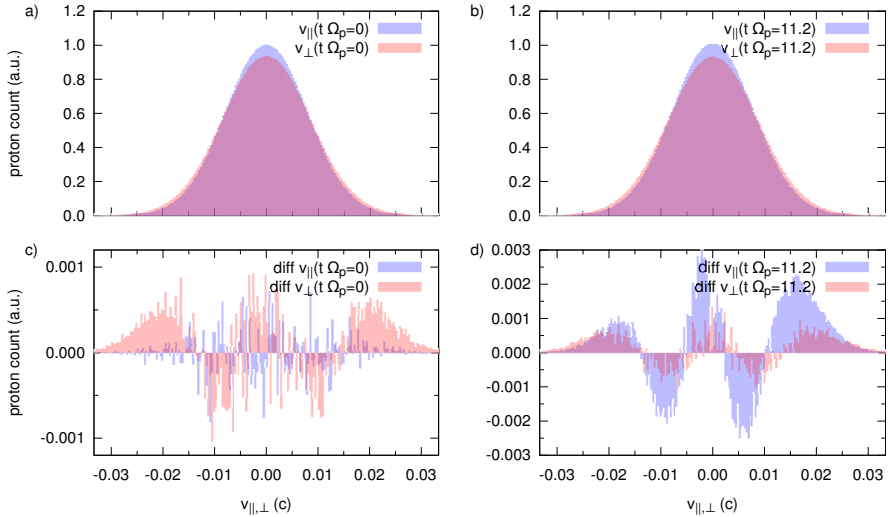}
	\caption{
		Distribution of the protons' parallel and perpendicular velocity components at the start (panel~a) and at the end (panel~b) of a simulation and deviations from a pure Gaussian distribution (panels~c and~d, respectively).
		Particle numbers are normalized to the number of particles in the bin around $v_\parallel = 0$ at the start of the simulation.
	}
	\label{fig:T_prot_components}
\end{figure}

Figure \ref{fig:T_prot_components} shows the protons' parallel and perpendicular velocity distributions at the start and at the end of one of the simulations of set $\mathrm{A}_1$.
The spectra are shown in the top panels (a and b) and the differences between the actual data and a Gaussian fit to the data are presented in the bottom panels (c and d).
Figure \ref{fig:T_prot_components} c) depicts the influence of the initial particle boost on the velocity spectra: while the deviations from a pure Gaussian are random in $v_{\parallel,\mathrm{p}}$, the deviations in $v_{\perp,\mathrm{p}}$ are systematic and symmetric around $v_{\perp,\mathrm{p}}=0$.
Excess particles can be found both at high and very low perpendicular speeds, whereas fewer particles can be found in the intermediate range.

At the end of the simulation, the perturbations in $v_{\perp,\mathrm{p}}$ have not changed in shape, but the perturbations in $v_{\parallel,\mathrm{p}}$ have (Fig. \ref{fig:T_prot_components} d).
Again, more particles can be found at high and very low speeds, and fewer particles are present in the intermediate range.
But unlike the deviations in $v_{\perp,\mathrm{p}}$, those in $v_{\parallel,\mathrm{p}}$ are not symmetric.
Deviations from the Gaussian distribution are stronger in the direction parallel to $\vec{B}$ and less pronounced in the direction anti-parallel to $\vec{B}$.

\begin{figure}[h!tb]
	\centering
	\includegraphics[width=0.8\linewidth]{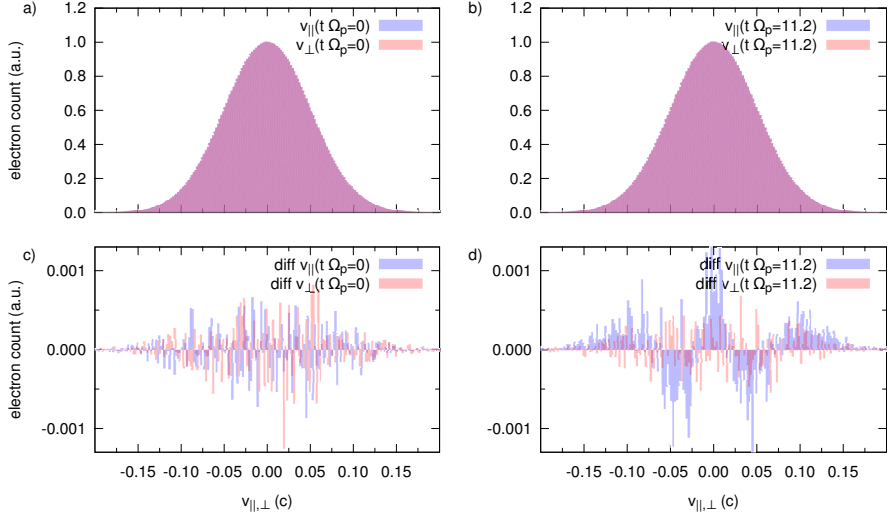}
	\caption{
		Distribution of the electrons' parallel and perpendicular velocity components at the start (panel~a) and at the end (panel~b) of a simulation and deviations from a pure Gaussian distribution (panels~c and~d).
		Particle numbers are normalized to the number of particles in the bin around $v_\parallel = 0$ at the start of the simulation.
	}
	\label{fig:T_elec_components}
\end{figure}

Similar figures showing the distribution of velocity components of the electrons are depicted in Fig. \ref{fig:T_elec_components}.
Again, we show the distributions of parallel and perpendicular velocity components of the electrons at the start and at the end of a simulation panels a) and b).
Panels c) and d) show the deviations of the actual data from Gaussian distributions.
These plots prove that the electrons' velocity spectra hardly diverge from a thermal distribution at the start of the simulation (panels a and c).
At the end of the simulation the spectrum of $v_{\parallel,\mathrm{e}}$ deviates from a Gaussian distribution similarly to $v_{\parallel,\mathrm{p}}$, but with a less prominent asymmetry (panel d).

We expect to find net momentum when summing up the momenta of all particles in the simulations.
Unfortunately, our analysis has shown that numerical fluctuations seem to dominate the total momentum of the particles, so that an effect of the decaying waves on the momentum of the whole particle population cannot be verified.

\section{Energy densities and damping rates obtained from set A}
\label{app:setA}

Table \ref{tab:damping_rate_A} contains the theoretical damping rates according to Eqs. (\ref{eq:warm_disp}) and (\ref{eq:warm_disp_maxwell}), as obtained with the parameters given in Table \ref{tab:setup_phys} and the temperatures discussed in Sect. \ref{sec:temperature}.
The mesured damping rates from set A (three-dimensional simulations) is also included.
Note that damping rates for waves with odd (even) $k_\mathrm{num}$ have to be compared to the theoretical values for $T_\mathrm{p} / T_\mathrm{e} = 1.10$ ($1.13$).

\begin{figure}[h!tb]
	\centering
	\includegraphics[width=\linewidth]{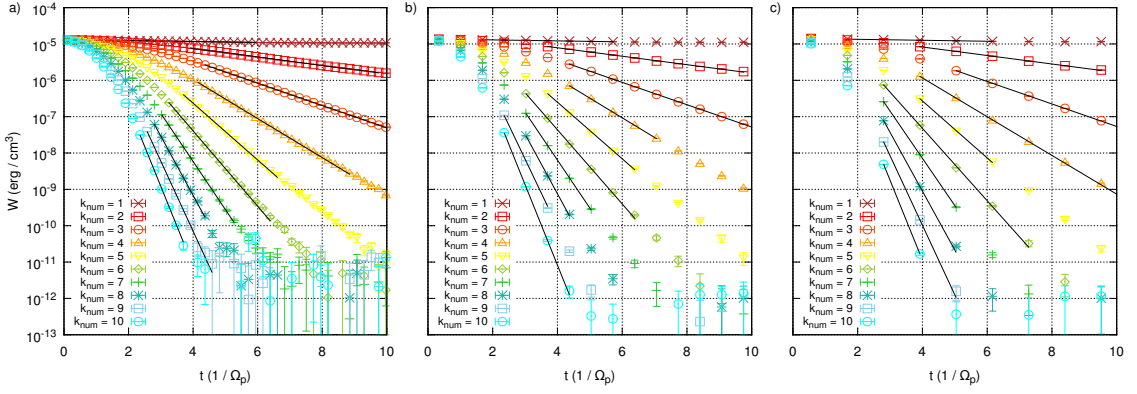}
	\caption{
		Energy density of ten waves obtained in intervals of $t_\mathrm{int} = 1000 \, \Delta t$ (panel a), $t_\mathrm{int} = 3000 \, \Delta t$ (panel b) and $t_\mathrm{int} = 5000 \, \Delta t$ (panel c) from the data of simulations of set A.
		Black lines indicate exponential fits to the data which yield the damping rates for the different waves.
	}
	\label{fig:energy_A}
\end{figure}

The average energy densities $W$ (see Sect. \ref{sec:damping_rate_sim} for details) of ten different waves in simulations $\mathrm{A}_1$ (odd $k_\mathrm{num}$) and $\mathrm{A}_2$ (even $k_\mathrm{num}$) are plotted over time in Fig. \ref{fig:energy_A}.
The three panels represent the data obtained from the simulations using the method described in Sect. \ref{sec:damping_rate} using intervals of length $t_\mathrm{int} = \{1000, \, 3000, \, 5000\} \, \Delta t$,  with $\Delta t$ being the length of one time step.
While shorter $t_\mathrm{int}$ yield higher temporal resolution (and more data points for the fit to be applied to), a more representative average of the waves' energies can be obtained during longer intervals, thus averaging out short time fluctuations.
However, the overall behavior is the same for all $t_\mathrm{int}$: Damping starts slowly (no exponential decay of energy), then develops the characteristic exponential slope representing the damping rate of each wave and finally cuts off when the energy of the wave is of the order of the background noise.

The background noise level $W_\mathrm{noise}$ seems to depend on the interval length, since $W_\mathrm{noise} \sim 10^{-11} \, \mathrm{erg} / \mathrm{cm}^3$ for $t_\mathrm{int} = 1000 \, \Delta t$ and $W_\mathrm{noise} \sim 10^{-12} \, \mathrm{erg} / \mathrm{cm}^3$ for $t_\mathrm{int} = \{3000, 5000\} \, \Delta t$.
Note that energy densities below $10^{-13} \, \mathrm{erg} \, / \, \mathrm{cm}^3$ are cut off, which is why there seem to be gaps in some data sets.
The measured energy density curves appear to deviate from the expected exponential decay already above the noise limit.
Most of the exponential fits have therefore been cut off at higher energy densities (see black lines in Fig. \ref{fig:energy_A}).
A reason for the unexpected deviation from the exponential decay could be the changing plasma temperature later on in the simulation.

\begin{table*}[htb]
	\centering
	\resizebox{\linewidth}{!}{%
	\begin{tabular}{c c c c c c}
		\hline
		\noalign{\smallskip}
		$k_\mathrm{num}$ & \multicolumn{4}{c}{$|\Gamma| \, (\Omega_\mathrm{p})$}\\
		& theory: & theory: & simulation: & simulation: & simulation: \\
		& $T_\mathrm{p} / T_\mathrm{e} = 1.10$ & $T_\mathrm{p} / T_\mathrm{e} = 1.13$ & $t_\mathrm{int} = 1000 \, \Delta t$ & $t_\mathrm{int} = 3000 \, \Delta t$ & $t_\mathrm{int} = 5000 \, \Delta t$ \\
		& $T_{\perp, \mathrm{p}} / T_{\parallel, \mathrm{p}} = 1.15$ & $T_{\perp, \mathrm{p}} / T_{\parallel, \mathrm{p}} = 1.20$ & & & \\
		\noalign{\smallskip}
		\hline
		\noalign{\smallskip}
		1 & $3.599 \cdot 10^{-4}$ & $3.548 \cdot 10^{-4}$ & $(1.5502 \pm 0.0018) \cdot 10^{-2}$ & $(1.555 \pm 0.003) \cdot 10^{-2}$ & $(1.549 \pm 0.002) \cdot 10^{-2}$ \\
		\noalign{\smallskip}
		2 & $0.1277$ & $0.1268$ & $0.13414 \pm 0.00004$ & $0.13383 \pm 0.00006$ & $0.13446 \pm 0.00007$ \\
		\noalign{\smallskip}
		3 & $0.3617$ & $0.3586$ & $0.35413 \pm 0.00011$ & $0.35392 \pm 0.00017$ & $0.3561 \pm 0.0003$ \\
		\noalign{\smallskip}
		4 & $0.6234$ & $0.6180$ & $0.6183 \pm 0.0004$ & $0.6224 \pm 0.0007$ & $0.6131 \pm 0.0005$ \\
		\noalign{\smallskip}
		5 & $0.9002$ & $0.8926$ & $0.8989 \pm 0.0010$ & $0.8958 \pm 0.0013$ & $0.9017 \pm 0.0018$ \\
		\noalign{\smallskip}
		6 & $1.189$ & $1.179$ & $1.1930 \pm 0.0016$ & $1.1744 \pm 0.0009$ & $1.1464 \pm 0.0008$ \\
		\noalign{\smallskip}
		7 & $1.486$ & $1.475$ & $1.523 \pm 0.004$ & $1.519 \pm 0.003$ & $1.4960 \pm 0.0017$ \\
		\noalign{\smallskip}
		8 & $1.792$ & $1.779$ & $1.878 \pm 0.008$ & $1.885 \pm 0.005$ & $1.863 \pm 0.004$ \\
		\noalign{\smallskip}
		9 & $2.105$ & $2.091$ & $2.218 \pm 0.015$ & $2.146 \pm 0.003$ & $2.202 \pm 0.012$ \\
		\noalign{\smallskip}
		10 & $2.425$ & $2.409$ & $2.524 \pm 0.017$ & $2.548 \pm 0.007$ & $2.53 \pm 0.04$ \\
		\noalign{\smallskip}
		\hline
	\end{tabular}
	}
	\caption{Theoretical and measured damping rates for ten different waves. Theoretical rates are given for two temperature settings, matching the setups used in simulations $\mathrm{A}_1$ and $\mathrm{A}_2$ to obtain the damping rates at odd and even numbered $k_\mathrm{num}$, respectively.}
	\label{tab:damping_rate_A}
\end{table*}

\section{Weakly damped wave}
\label{app:weak_damping}

\begin{figure}[h]
	\centering
	\includegraphics[width=0.6\linewidth]{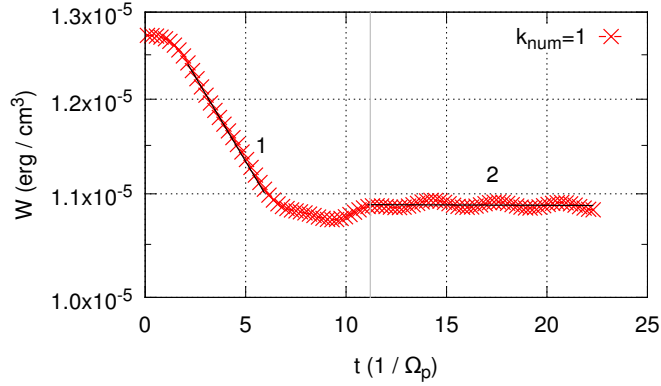}
	\caption{
		Energy density $W$ of an excited wave with $k_\mathrm{num} = 1$ in a rerun of a simulation of type $\mathrm{A}_1$.
		The gray line marks the end of the regular simulation, the black line labeled with '1' represents the exponential fit which yields $\Gamma (k_\mathrm{num} = 1)$ for $t_\mathrm{int} = 1000 \, \Delta t$ in Table \ref{tab:damping_rate_A} in Appendix \ref{app:setA}.
		The rerun simulation contains twice as many time steps as the original run.
		Thus, an additional exponential fit can be obtained at later times, as indicated by the black line labeled with '2'.
		Note that the y-axis is in logarithmic scale.
	}
	\label{fig:gamma1}
\end{figure}

One peculiarity in the data shown in Table \ref{tab:damping_rate_A} in Appendix \ref{app:setA} is the measured damping rate for $k_\mathrm{num} = 1$, which is almost two orders of magnitude above the theoretical expectations.
We therefore examine the behavior of the wave with $k_\mathrm{num} = 1$ more closely in a rerun of one simulation of set $\mathrm{A}_1$.
The energy density $W$ is plotted over time $t$ in Fig. \ref{fig:gamma1}, where $t_\mathrm{int} = 1000 \, \Delta t$ is used.
The rerun uses the same physical and numerical parameters as the original simulation (see Tables \ref{tab:setup_phys} and \ref{tab:setup_num} in Sect. \ref{sec:setup}), but employs twice as many time steps, thus doubling the physical time simulated.

The end of the regular simulation of type $\mathrm{A}_1$ is indicated by a gray line in Fig. \ref{fig:gamma1}.
At the start of the simulation the energy density decreases exponentially, but seems to level out towards the end of the regular simulation time, with the energy density still being way above the noise limit ($W_\mathrm{noise} \approx 10^{-11} \, \mathrm{erg}/\mathrm{cm}^3$).
The fit which lead to $\Gamma (k_\mathrm{num} = 1)$ for $t_\mathrm{int} = 1000 \, \Delta t$ in Table \ref{tab:damping_rate_A}  in Appendix \ref{app:setA} is shown as a black line labeled with '1'.
In the second half of the simulation the energy density stays at a rather constant level, fluctuating only slightly.
Applying another exponential fit to this regime (labeled with '2' in Fig. \ref{fig:gamma1}) yields a different damping rate: $\Gamma^\mathrm{rerun} (k_\mathrm{num} = 1) \cdot \Omega_\mathrm{p} = -(3 \pm 5) \cdot 10^{-5}$.
Since only one simulation has been carried out, no standard error can be calculated and only the statistical error from the fit is given.

Although the newly obtained damping rate is closer to the theoretical expectations, it is not very reliable due to the large error.
Additionally, the runtime of the simulation has to be twice as long as originally planned, thus doubling the computational effort of the simulation.
The reason for the first exponential decay of the wave, visible between $t \, \Omega_\mathrm{p} = 2$ and $t \, \Omega_\mathrm{p} = 6$ in Fig. \ref{fig:gamma1}, is unclear.
However, no further investigations have been carried out, since the regime of very weak damping ($|\Gamma| \ll \Omega_\mathrm{p}$) is not in the focus of this study.

\section{Energy densities and damping rates obtained from set B}
\label{app:setB}

Table \ref{tab:damping_rate_B} contains theoretical and measured damping rates for the two-dimensional simulations of set B.
Again, we give the theoretical values for two different proton temperatures and temperature anisotropies, as discussed in in Sect. \ref{sec:temperature}.
Errors of measured data are standard errors.
No damping rates are given for the $k_\mathrm{num} = 1$, as announced at the end of Sect. \ref{sec:damping_rate_sim}.

\begin{figure}[htb]
	\centering
	\includegraphics[width=\linewidth]{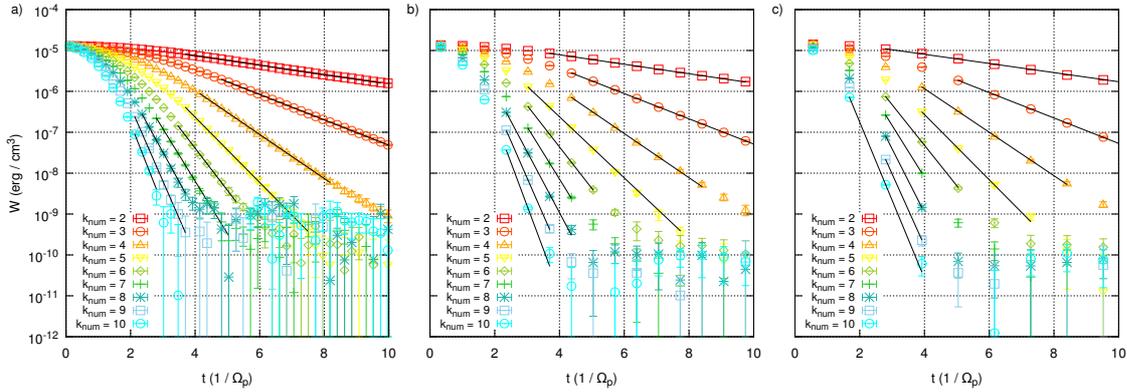}
	\caption{Energy density of nine waves obtained in intervals of $t_\mathrm{int} = 1000 \, \Delta t$ (panel a), $t_\mathrm{int} = 3000 \, \Delta t$ (panel b) and $t_\mathrm{int} = 5000 \, \Delta t$ (panel c) from the data of simulations of set B. Black lines indicate exponential fits to the data which yield the damping rates for the different waves.}
	\label{fig:energy_B}
\end{figure}

The time evolution of the energy density $W$ is presented in Fig. \ref{fig:energy_B} for three different interval lengths $t_\mathrm{int}$.
Note that the noise level is around $W_\mathrm{noise} \sim 10^{-10} \, \mathrm{erg} / \mathrm{cm}^{3}$, compared to $W_\mathrm{noise} \sim 10^{-12} \, \mathrm{erg} / \mathrm{cm}^{3}$ in the three-dimensional simulations (see Fig. \ref{fig:energy_B}).

\begin{table*}[ht]
	\centering
	\resizebox{\linewidth}{!}{
	\begin{tabular}{c c c c c c}
		\hline
		\noalign{\smallskip}
		$k_\mathrm{num}$ & \multicolumn{4}{c}{$|\Gamma| \, (\Omega_\mathrm{p})$}\\
		& theory: & theory: & simulation: & simulation: & simulation: \\
		& $T_\mathrm{p} / T_\mathrm{e} = 1.10$ & $T_\mathrm{p} / T_\mathrm{e} = 1.13$ & $t_\mathrm{int} = 1000 \, \Delta t$ & $t_\mathrm{int} = 3000 \, \Delta t$ & $t_\mathrm{int} = 5000 \, \Delta t$ \\
		& $T_{\perp, \mathrm{p}} / T_{\parallel, \mathrm{p}} = 1.15$ & $T_{\perp, \mathrm{p}} / T_{\parallel, \mathrm{p}} = 1.20$ & & & \\
		\noalign{\smallskip}
		\hline
		\noalign{\smallskip}
		2 & $0.1277$ & $0.1268$ & $0.1335 \pm 0.0002$ & $0.1334 \pm 0.0003$ & $0.1287 \pm 0.0003$ \\
		\noalign{\smallskip}
		3 & $0.3617$ & $0.3586$ & $0.3574 \pm 0.0010$ & $0.3555 \pm 0.0013$ & $0.3577 \pm 0.0018$ \\
		\noalign{\smallskip}
		4 & $0.6234$ & $0.6180$ & $0.620 \pm 0.003$ & $0.622 \pm 0.004$ & $0.612 \pm 0.004$ \\
		\noalign{\smallskip}
		5 & $0.9002$ & $0.8926$ & $0.920 \pm 0.005$ & $0.860 \pm 0.003$ & $0.918 \pm 0.010$ \\
		\noalign{\smallskip}
		6 & $1.189$ & $1.179$ & $1.186 \pm 0.018$ & $1.168 \pm 0.009$ & $1.141 \pm 0.008$ \\
		\noalign{\smallskip}
		7 & $1.486$ & $1.475$ & $1.476 \pm 0.016$ & $1.471 \pm 0.018$ & $1.458 \pm 0.015$ \\
		\noalign{\smallskip}
		8 & $1.792$ & $1.779$ & $1.66 \pm 0.04$ & $1.715 \pm 0.013$ & $1.793 \pm 0.043$ \\
		\noalign{\smallskip}
		9 & $2.105$ & $2.091$ & $2.08 \pm 0.04$ & $2.08 \pm 0.03$ & $2.02 \pm 0.13$ \\
		\noalign{\smallskip}
		10 & $2.425$ & $2.409$ & $2.31 \pm 0.09$ & $2.43 \pm 0.05$ & $2.197 \pm 0.013$ \\
		\noalign{\smallskip}
		\hline
	\end{tabular}
	}
	\caption{Theoretical and measured damping rates for nine different waves. Theoretical rates are given for two temperature settings, matching the setups used in simulations $\mathrm{B}_1$ and $\mathrm{B}_2$ to obtain the damping rates at odd and even numbered $k_\mathrm{num}$, respectively.}
	\label{tab:damping_rate_B}
\end{table*}

\section{Energy densities and damping rates obtained from set C}
\label{app:setC}

Table \ref{tab:damping_rate_C} contains theoretical and measured damping rates from the one-dimensional simulations of set C.
As before, $k_\mathrm{num} = 1$ is omitted and theoretical damping rates are computed for two different temperature configurations.

\begin{figure}[htb]
	\centering
	\includegraphics[width=\linewidth]{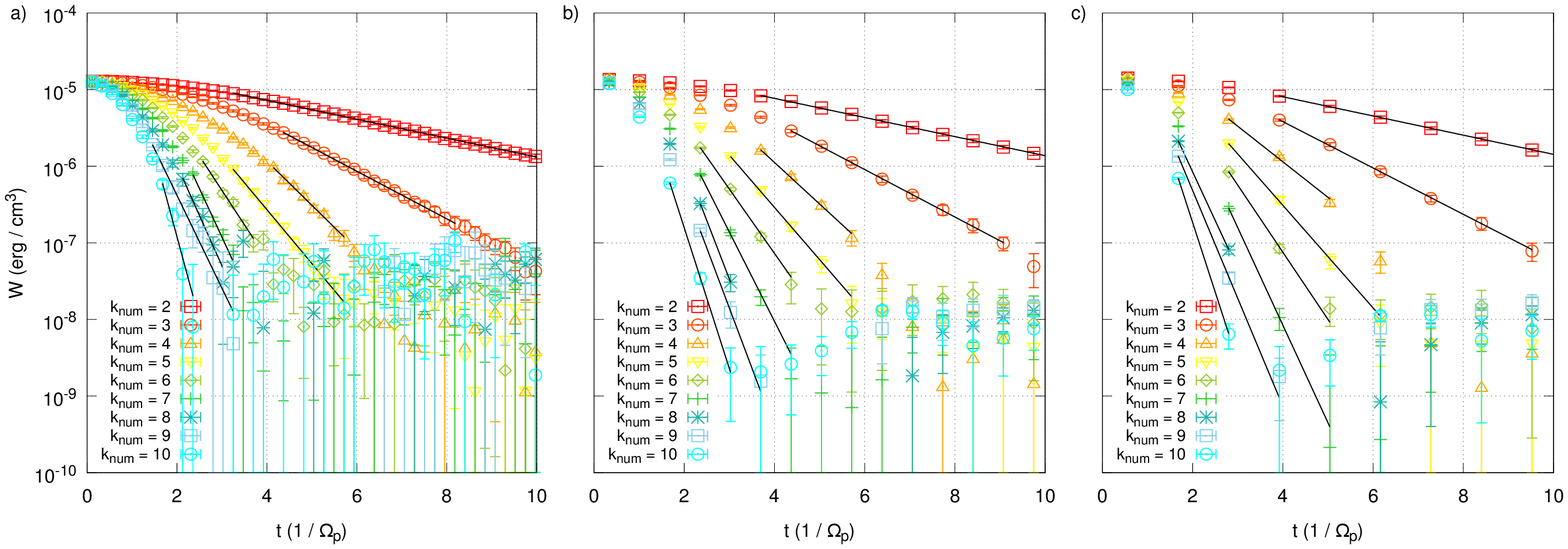}
	\caption{Energy density of nine waves obtained in intervals of $t_\mathrm{int} = 1000 \, \Delta t$ (panel a), $t_\mathrm{int} = 3000 \, \Delta t$ (panel b) and $t_\mathrm{int} = 5000 \, \Delta t$ (panel c) from the data of simulations of set C. Black lines indicate exponential fits to the data which yield the damping rates for the different waves.}
	\label{fig:energy_C}
\end{figure}

Figure \ref{fig:energy_C} shows the time evolution of the energy density $W$ for the excited waves in the simulations of set C.
As stated at the end of Sect. \ref{sec:1D} it becomes obvious that the one-dimensional setup leads to a drastic increase of the background noise $W_\mathrm{noise}$.
With $W_\mathrm{noise} \sim 10^{-8} \, \mathrm{erg} / \mathrm{cm}^3$ and the initial amplitude of the waves $W(t=0) \sim 10^{-5} \, \mathrm{erg} / \mathrm{cm}^3$, only three orders of magnitude remain over which the damping rate can be determined.
However, with the sub-exponential decay in the early time during the simulation, the exponential fits can only be applied over two orders of magnitude.
This explains the data quality, which is considerably worse than in the two- and three-dimensional simulations.

\begin{table*}[ht]
	\centering
	\resizebox{\linewidth}{!}{%
	\begin{tabular}{c c c c c c}
		\hline
		\noalign{\smallskip}
		$k_\mathrm{num}$ & \multicolumn{4}{c}{$|\Gamma| \, (\Omega_\mathrm{p})$}\\
		& theory: & theory: & simulation: & simulation: & simulation: \\
		& $T_\mathrm{p} / T_\mathrm{e} = 1.10$ & $T_\mathrm{p} / T_\mathrm{e} = 1.13$ & $t_\mathrm{int} = 1000 \, \Delta t$ & $t_\mathrm{int} = 3000 \, \Delta t$ & $t_\mathrm{int} = 5000 \, \Delta t$ \\
		& $T_{\perp, \mathrm{p}} / T_{\parallel, \mathrm{p}} = 1.15$ & $T_{\perp, \mathrm{p}} / T_{\parallel, \mathrm{p}} = 1.20$ & & & \\
		\noalign{\smallskip}
		\hline
		\noalign{\smallskip}
		2 & $0.1277$ & $0.1268$ & $0.1422 \pm 0.0014$ & $0.143 \pm 0.003$ & $0.144 \pm 0.003$ \\
		\noalign{\smallskip}
		3 & $0.3617$ & $0.3586$ & $0.357 \pm 0.010$ & $0.357 \pm 0.012$ & $0.348 \pm 0.010$ \\
		\noalign{\smallskip}
		4 & $0.6234$ & $0.6180$ & $0.66 \pm 0.04$ & $0.62 \pm 0.03$ & $0.534 \pm 0.017$ \\
		\noalign{\smallskip}
		5 & $0.9002$ & $0.8926$ & $0.81 \pm 0.03$ & $0.78 \pm 0.03$ & $0.77 \pm 0.03$ \\
		\noalign{\smallskip}
		6 & $1.189$ & $1.179$ & $1.03 \pm 0.06$ & $0.97 \pm 0.03$ & $1.01 \pm 0.05$ \\
		\noalign{\smallskip}
		7 & $1.486$ & $1.475$ & $1.46 \pm 0.10$ & $1.34 \pm 0.05$ & $1.47 \pm 0.15$ \\
		\noalign{\smallskip}
		8 & $1.792$ & $1.779$ & $1.5 \pm 0.2$ & $1.76 \pm 0.19$ & $1.46 \pm 0.05$ \\
		\noalign{\smallskip}
		9 & $2.105$ & $2.091$ & $1.40 \pm 0.10$ & $1.8 \pm 0.3$ & $1.62 \pm 0.07$ \\
		\noalign{\smallskip}
		10 & $2.425$ & $2.409$ & $2.5 \pm 0.5$ & $2.11 \pm 0.12$ & $2.09 \pm 0.16$ \\
		\noalign{\smallskip}
		\hline
	\end{tabular}
	}
	\caption{Theoretical and measured damping rates for nine different waves. Theoretical rates are given for two temperature settings, matching the setups used in simulations $\mathrm{C}_1$ and $\mathrm{C}_2$ to obtain the damping rates at odd and even numbered $k_\mathrm{num}$, respectively.}
	\label{tab:damping_rate_C}
\end{table*}

\clearpage


\end{document}